\documentclass[useAMS,usenatbib]{mn2e}
\usepackage{amssymb,amstext,amsmath,commath}
\usepackage{times,float,graphicx,paralist}
\usepackage{epsfig,epstopdf}
\DeclareGraphicsExtensions{.pdf,.ps,.eps,.png,.jpg,.mps} 
\thinspace
\begin{document}
\title[Satellite quenching in $z\sim1$ Galaxy Groups] {Efficient satellite quenching at $z\sim1$ from the GEEC2 spectroscopic survey of galaxy groups}
\author[Mok et al.] {Angus Mok$^{1}$, Michael L. Balogh$^{1}$, Sean L. McGee$^{2,6}$, David J. Wilman$^{3}$,
\newauthor Alexis Finoguenov$^{4,9}$, Masayuki Tanaka$^{5}$, Stefania Giodini$^{6}$, Richard G. Bower$^{2}$, 
\newauthor Jennifer L. Connelly$^{3}$, Annie Hou$^{7}$, John~S. Mulchaey$^{8}$, Laura C. Parker$^{7}$\\
$^{1}$Department of Physics and Astronomy, University of Waterloo, Waterloo, Ontario, N2L 3G1, Canada\\
$^{2}$Department of Physics, University of Durham, Durham, UK, DH1 3LE\\
$^{3}$Max--Planck--Institut f{\" u}r extraterrestrische Physik, Giessenbachstrasse 85748 Garching Germany\\
$^{4}$CSST, University of Maryland, Baltimore County, 1000 Hilltop Circle, Baltimore, MD 21250, USA\\
$^{5}$Institute for the Physics and Mathematics of the Universe, University of Tokyo, Kashiwa 2778582, Japan\\
$^{6}$Leiden Observatory, Leiden University, PO Box 9513, 2300 RA Leiden,the Netherlands\\
$^{7}$Department of Physics and Astronomy, McMaster University, Hamilton, Ontario, L8S 4M1 Canada\\
$^{8}$Observatories of the Carnegie Institution, 813 Santa Barbara Street, Pasadena, California, USA\\
$^{9}$Department of Physics, University of Helsinki, Gustaf H\"allstr\"omin katu 2a, FI-00014 Finland\\
}
%
%
\def\etal{{ et al.\thinspace}}
\def\gtrsim{\mathrel{\raise0.35ex\hbox{$\scriptstyle >$}\kern-0.6em\lower0.40ex\hbox{{$\scriptstyle \sim$}}}}
\def\lesssim{\mathrel{\raise0.35ex\hbox{$\scriptstyle <$}\kern-0.6em\lower0.40ex\hbox{{$\scriptstyle \sim$}}}}
\def\Msun{\hbox{$\rm\thinspace M_{\odot}$}}
\def\kmsmpc{{\,\rm km\,s^{-1}Mpc^{-1}}}
\def\kms{km~s$^{-1}$}
\def\oii{[O{\sc ii}]\ }
\def\ewoii{$W_{\rm \circ}$(\oii)}
\def\ewhdelta{$W_{\rm \circ}$(\hdelta)}
\def\hdelta{H$\delta\ $}
%
%
\maketitle
\begin{abstract}
We present deep GMOS-S spectroscopy for 11 galaxy groups at $0.8<z<1.0$, for galaxies with $r_{AB}<24.75$. Our sample is highly complete ($>66\%$) for eight of the eleven groups. Using an optical-NIR colour-colour diagram, the galaxies in the sample were separated with a dust insensitive method into three categories: passive (red), star-forming (blue), and intermediate (green). The strongest environmental dependence is observed in the fraction of passive galaxies, which make up only $\sim20$ per cent of the field in the mass range $10^{10.3}<M_{\rm star}/M_\odot<10^{11.0}$, but are the dominant component of groups. If we assume that the properties of the field are similar to those of the `pre-accreted' population, the environment quenching efficiency ($\epsilon_\rho$) is defined as the fraction of field galaxies required to be quenched in order to match the observed red fraction inside groups. The efficiency obtained is $\sim 0.4$, similar to its value in intermediate-density environments locally. While green (intermediate) galaxies represent $\sim20$ per cent of the star-forming population in both the group and field, at all stellar masses, the average sSFR of the group population is lower by a factor of $\sim3$. The green population does not show strong \hdelta absorption that is characteristic of starburst galaxies. Finally, the high fraction of passive galaxies in groups, when combined with satellite accretion models, require that most accreted galaxies have been affected by their environment. Thus, any delay between accretion and the onset of truncation of star formation ($\tau$) must be $\lesssim2$ Gyr, shorter than the $3-7$ Gyr required to fit data at $z=0$. The relatively small fraction of intermediate galaxies requires that the actual quenching process occurs quickly, with an exponential decay timescale of $\tau_q\lesssim 1$ Gyr.
\end{abstract}
%
%
\section{Introduction}
\par
Studies of galaxy evolution have shown that the star formation density of the universe depends on epoch \citep[e.g.][]{1996ApJ...460L...1L}, stellar mass \citep[e.g.][]{2004MNRAS.351.1151B}, and environment \citep[e.g.][]{2006MNRAS.373..469B, 2011MNRAS.411..675S}. At $z\sim1$, star-forming galaxies continue to show a strong relationship between the specific star formation rate (sSFR) and stellar mass, with moderate scatter \citep{2007ApJ...660L..43N, 2012ApJ...754L..14S}. Galaxies that lie above this sequence tend to either host an active galactic nucleus (AGN) or a short-lived burst of star formation, while those below the relation may be cases where the galaxy's star formation is shutting down \citep{2012ApJ...754L..29W}.
\par
Groups are an important test of the effects of environment on galaxy evolution, as the majority of galaxies are found inside groups \citep{2004MNRAS.355..769E} and they are a crucial link in the hierarchical growth of structures from individual galaxies to dense clusters. Galaxy groups have been studied in the context of general redshift surveys, for example, CNOC2 \citep{2001ApJ...563..736C, 2009ApJ...704..564F}, DEEP2 \citep{2012ApJ...751...50G}, and zCOSMOS \citep{2012ApJ...753..121K}, including imaging-only photometric z work \citep[e.g.][]{2012A&A...538A.104G, 2012ApJ...749..150L, 2012ApJ...746...95L}. Another approach is to study individual groups in more detail and with highly complete spectroscopy \citep[e.g.][]{2009MNRAS.399..715J, 2009A&A...508.1173P, 2011MNRAS.412.2303B}.
\par
Observations have also demonstrated that there already exists a strong correlation between galaxy environment and the fraction of passive galaxies at $z\sim 1$ \citep[e.g.][]{2006MNRAS.370..198C, 2010A&A...523A..13P}. In general, the fraction of quiescent galaxies is always higher in denser and more massive systems \citep{2008MNRAS.387...79V, 2011ApJ...742..125G, 2012ApJ...746..188M, 2012PASJ...64...22T, 2012A&A...539A..55P}, except where the galaxy merger rate may be exceptionally high \citep[e.g.][]{2011MNRAS.413..177G, 2011MNRAS.411..675S}. At $z>0.8$, there is evidence that, at least in some cases, the SFR of the star-forming population in locally-dense environments may in fact be higher than in the field \citep{2008MNRAS.383.1058C, 2007A&A...468...33E, 2011MNRAS.411..675S, 2011MNRAS.411.1869L}, while other groups have not found this enhancement \citep[e.g.][]{2011ApJ...735...53P}.
\par
It is perhaps most natural to interpret these observations in the context of the halo model, in which the properties of the host dark matter halo completely determine the galaxy's properties and where the evolution of satellite galaxies can be distinguished from that of central galaxies. In this context, the radial gradients of the quiescent fraction observed in groups and cluster \citep[e.g.][]{2000ApJ...540..113B, 2001ApJ...547..609E, 2012A&A...539A..55P} reflect the satellites' accretion history \citep[e.g.][]{2012MNRAS.419.3167S, 2012MNRAS.423.1277D}.
\par
An empirical model of galaxy evolution that accounts for most of these trends was put forward by \citet{2010ApJ...721..193P, 2012ApJ...757....4P}. In this model, galaxy evolution is driven by two ``quenching rates'', one that depends, perhaps indirectly, on stellar mass and another that is related to environment (local density). While the efficiency of environmental quenching in this model is roughly independent of stellar mass and redshift, its effects are most prominent amongst low-mass galaxies, $M<10^{10.5}M_\odot$, for which mass-quenching is inefficient. By trying to match the full shape of the observed SFR distribution at $z=0$, \citet{2012arXiv1206.3571W} recently proposed that this environmental dependence must take the form of a rapid ($\tau<1$ Gyr) satellite quenching timescale, but preceded by a long delay of $2-4$ Gyr.
\par
Both of these models make predictions about the environmental dependence of galaxies at $z=1$, which are poorly constrained by existing observations, in part because the effects are expected to be most visible in the low-mass star-forming population. In particular, such a long delay before a satellite begins to respond to its changed environment means that trends with environment should be much weaker at $z\sim 1$, since at that epoch most galaxies would not have spent a long enough time as a satellite \citep{2009MNRAS.400..937M}.
\par
For all of the reasons listed above, the study of the universe at $0.8<z<1$ is particularly important. Compared to $z\sim0$, this epoch has high in-fall rates into groups, and the galaxies themselves have high star formation rates and gas fractions, thus presenting the best opportunity to observe any transitional objects \citep{2009MNRAS.400..937M}.
\par
This paper will present the data collected as part of the Group Environment Evolution Collaboration 2 (GEEC2) study \citep{2011MNRAS.412.2303B}. Our survey selects a sample of $\sim20$ robustly-identified groups with extended X-ray emissions \citep{2007ApJS..172..182F, 2011ApJ...742..125G} from the COSMOS survey fields \citep{2007ApJS..172....1S}. Then, we seek to obtain deep, highly complete spectroscopy for their member galaxies, focusing on low mass galaxies down to $10^9 \mathrm{M}_\odot$, where environmental effects may be most prominent. This approach of surveying a smaller number of well-sampled groups has been previously applied at $z\sim0.5$ with great success \citep[e.g.][]{2005MNRAS.358...71W, 2007MNRAS.374.1169B, 2008MNRAS.387.1605M, 2009ApJ...692..298W, 2009MNRAS.398..754B, 2009ApJ...704..564F, 2012ApJ...756..139C}.
\par
Analysis of low-mass groups at $z\sim1.0$ has revealed a distinct population of intermediate-colour (green) galaxies \citep{2011MNRAS.412.2303B}, which may indicate the presence of a transition population. These green galaxies have morphologies half-way between the blue-cloud and red-sequence galaxies and do not appear to be exceptionally dusty. Inside the group environment, they may be evidence for the impact of environment on star formation.  The sample has since been expanded with additional GMOS observations, allowing us to improve the statistical significance of these trends, and study their SFR and stellar mass properties.
\par
The structure of the paper will be as follows. In \S~2, we present a discussion of the updated GEEC2 sample with new observations from the 11A semester and the updated data reduction process. In \S~3, the main analysis of the data will be presented, including our separation of the sample into sub-populations using their colour, the determination of completeness weightings, and the calculation of star-formation rates. In \S~4, the results are presented, including stellar mass functions and the comparison of sSFR and \hdelta strengths for different sub-populations and environment. Finally, \S~5 will include a discussion of how satellite accretion models and the red/green/blue fractions in our sample constrain quenching models and the effects of environment on the group population. A companion paper (Hou et al., in prep) considers how the galaxy population in these groups depend on the group dynamics.
\par
Throughout the paper, a cosmology with $\Omega_m=0.272$, $\Omega_\Lambda=1-\Omega_m$ and $h=H_\circ/\left(100 \mbox{km}/\mbox{s}/\mbox{Mpc}\right)=0.702$ was assumed \citep{2011ApJS..192...18K}. All magnitudes are in the AB system and star formation rates are given for the Chabrier IMF.
%
%
\section{Observations}
\par
A more complete discussion and explanation of the GEEC2 survey can be found in \citet{2011MNRAS.412.2303B}, which presents results from the first semester of Gemini data. The survey is designed to cover 20 groups at $0.8<z<1$, requiring $\sim160$ hours of Gemini time over three years. So far, a total of 82.5 hours of data have been collected from Semesters 10A and 11A, covering approximately 11 groups. The following is a summary of the most important details, including an update of the sample to include observations from Semester 11A.
\subsection{Galaxy Group Selection}
\par
The GEEC2 survey targets galaxy groups identified within the COSMOS field \citep{2007ApJS..172....1S}, based on analysis of the deep X-ray data \citep{2007ApJS..172..182F, 2009ApJ...704..564F, 2010MNRAS.403.2063F}. Identification of group redshifts is done using all available spectroscopy, including but not limited to, the 10K zCOSMOS survey. 
\par
We considered all groups presented in the \citet{2011ApJ...742..125G, 2012ApJ...757....2G} catalogue that lie within the redshift range $0.8<z<1$, have at least three spectroscopically-determined members, and were classified as either category 1 (good detection, with a well-determined X-ray centre) or category 2 (secure detection, but with an unreliable X-ray centre). There are 21 groups satisfying this selection. Of these, we have given preference to the lowest-mass, highest redshift systems. The X-ray selected groups in this paper are a subset of the groups in the \citet{2012ApJ...753..121K} spectroscopic catalogue.
\par
Over two semesters with Gemini we observed ten of these groups, and from these data discovered an additional serendipitous group in the same redshift range (213a). Their properties are tabulated in Table~\ref{tab-gprop}.
\begin{table*}
\begin{minipage}{150mm}
	\begin{tabular}{|l|l|l|l|l|l|l|l|l|l|l|}
 	\hline
 	Group & RA & Dec & $z_\mathrm{mean}$ & $N_{mask}$ & $N_{mem}$ &
 $\sigma$ (km/s) & $R_\mathrm{rms}$ (Mpc) & Comp. ($\%$)
 & $M_{\rm dyn}(10^{13}M_\odot)$ \\
		\hline 
40 & 150.414 & 1.848 & 0.9713 & 2 & 15 & $690\pm110$ & $0.34\pm0.04$ & 64 & $11\pm4.7$ \\ 
71 & 150.369 & 1.999 & 0.8277 & 3 & 21 & $360\pm40$ & $0.34\pm0.03$ & 82 & $3.0\pm0.9$ \\ 
120 & 150.505 & 2.225 & 0.8358 & 3 & 31 & $480\pm60$ & $0.78\pm0.07$ & 69 & $12.7\pm4.2$ \\ 
121 & 150.161 & 2.137 & 0.8373 & 3 & 5 & $210\pm60$ & $0.09\pm0.01$ & 100 & $0.2\pm0.2$ \\ 
130 & 150.024 & 2.203 & 0.9374 & 3 & 34 & $600\pm70$ & $0.71\pm0.06$ & 76 & $17.8\pm5.7$ \\ 
134 & 149.65 & 2.209 & 0.9467 & 3 & 23 & $450\pm60$ & $0.97\pm0.07$ & 70 & $13.3\pm4.7$ \\ 
143 & 150.215 & 2.28 & 0.881 & 3 & 20 & $580\pm60$ & $0.23\pm0.03$ & 100 & $5.2\pm1.7$ \\ 
150 & 149.983 & 2.317 & 0.9334 & 4 & 25 & $300\pm40$ & $0.89\pm0.08$ & 75 & $5.5\pm2$ \\ 
161 & 149.953 & 2.342 & 0.944 & 4 & 8 & $170\pm30$ & $0.53\pm0.12$ & 70 & $1.0\pm0.5$ \\ 
213 & 150.41 & 2.512 & 0.879 & 2 & 9 & $260\pm100$ & $0.84\pm0.13$ & 52 & $3.8\pm3.4$ \\ 
213a & 150.428 & 2.505 & 0.9256 & 2 & 8 & $110\pm30$ & $0.62\pm0.09$ & 40 & $0.4\pm0.3$ \\ 
 	\hline
 	\end{tabular}
	\caption{Properties of the eleven galaxy groups observed with GMOS in semesters 10A and 11A. The position, median redshift $z_{\rm med}$, rest-frame velocity dispersion, and the number of group members are determined from our GMOS spectroscopy (combined with available zCOSMOS 10k data), as described in the text. The number of GMOS masks observed in each field is given by $N_{\rm mask}$; note that groups 150 and 161 lie within the same field, as do groups 213 and 213a (a serendipitous discovery in the background). $R_{\rm rms}$ is the rms projected distance of all group members from the centre. The penultimate column lists the spectroscopic completeness of the group, where we list the percentage of candidates within $R_{\rm rms}$ of the group centre for which spectroscopic redshift was obtained, where a candidate is a galaxy with a photometric redshift consistent with the group redshift at the 2$\sigma$ level (see \S~\ref{sec-zsuc}). Finally we list the dynamical mass of the group, calculated from $\sigma$ and $R_{\rm rms}$ following Equation~\ref{eqn-mdyn}.}
	\label{tab-gprop}
\end{minipage}
\end{table*}
\subsection{Gemini Observations} \label{sec-gemini}
\subsubsection{Spectroscopic Target Selection} \label{sec-specsel}
\par
Targets were selected from the COSMOS photometric catalogue \citep{2007ApJS..172...99C}. All optical colours and magnitudes are computed within a 3\arcsec\ diameter aperture. A crucial part of our strategy is the use of photometric redshifts to select potential group members. Precise photometric redshifts are derived from the photometry of 30 broad, intermediate and narrow-band filters \citep{2009ApJ...690.1236I}, using a template-fitting technique, and calibrated based on large spectroscopic samples. Even for the faintest galaxies in our sample, most of the objects have photometric redshifts determined to better than 10 per cent.
\par
We give highest priority to galaxies that have $21.5<r<24.75$, and a redshift within 2$\sigma_{zphot}$ of the estimated group redshift, where $\sigma_{zphot}$ is the 68 per cent confidence level on the photometric redshift \footnote{At our redshift of interest, the average uncertainty on $z_{\rm phot}$ increases from $\sigma_{zphot}\sim 0.007$ for the brightest galaxies to $\sigma_{zphot}\sim 0.04$ for those at our limit of $r=24.75$. Even for these faintest objects, 90 per cent of the galaxies have $z_{\rm phot}$ uncertainties of less than $\sigma_{zphot}<0.07$.}. Secondary priority slits are allocated to galaxies with $15<r<24.75$ and $0.7<z_{\rm phot}<1.5$. 
\par
We designed $3-4$ GMOS masks on each target, using the {\sc gmmps} software. On each mask we are able to allocate $40-50$ slits. Typically, well over half the slits on the first mask are allocated to our top priority objects. This fraction decreases on subsequent masks, as we use up these targets. In most cases, with three masks we are able to target at least 40 of these high priority galaxies. Many masks for a given target use some of the same alignment stars and are always at the same position angle; thus a small fraction of the CCD area is unusable regardless of the number of masks obtained.
\subsubsection{11A Gemini Observations}
\par
Our 10A observations were described in \citet{2011MNRAS.412.2303B}. In 11A we obtained an additional 32.5 hours of data from Gemini South, in the Band 1 queue. This returned science data of 2 hours on-source exposure, for 12 masks in 4 groups. All science observations were obtained in clear conditions with seeing 0.8 arcsec or better in $i$.
\par
The spectroscopic setup was almost the same as in \citet{2011MNRAS.412.2303B}. We used the nod \& shuffle mode \citep{2001PASP..113..197G}, observing the galaxy in both nodded positions. Slits were 1 arcsec wide, and we use the R600 grism with the OG515 order blocking filter. One difference from the 10A observations is the detector was binned by $4$ in the spectral direction, rather than $2$. This yielded a dispersion of $1.86$ \AA/pix, and a spectral resolution, limited by the slit width, of $\sim 12.8$ \AA.
\subsubsection{Data reduction}\label{sec-dr}
\par
All data are reduced in {\sc iraf}, using the {\sc gemini} packages with minor modifications as described in \citet{2011MNRAS.412.2303B}. Redshifts are measured by adapting the {\sc zspec} software, kindly provided by R. Yan, used by the DEEP2 redshift survey \citep{2012arXiv1203.3192N}. This performs a cross-correlation on the 1D extracted spectra, using linear combinations of template spectra. The corresponding variance vectors are used to weight the cross-correlation. Finally, redshifts are adjusted to the local standard of rest using the {\sc iraf} task {\sc rvcorrect}, though this correction is negligible.
\par
We adopt a simple, four-class method to quantify our redshift quality. Quality class 4 is assigned to galaxies with certain redshifts. Generally this is reserved for galaxies with multiple, robust features. Quality class 3 are also very reliable redshifts, and we expect most of them to be correct. These include galaxies with a good match to Ca H\&K for example, but no obvious corroborating feature.
\par
Note our spectral resolution was a factor of two lower in our 11A observations, and this meant that it was generally not possible to resolve the \oii doublet. Thus, single emission lines are generally assumed to be \oii and given quality class 3. This is, in part, due to our prior photo-z data for these objects. We would also use this in cases where H\&K are detected in a region of telluric absorption, but there is at least one other likely match to an absorption feature. We take particular care not to assign class 3 or 4 to a galaxy for which H\&K are the only identifiable features, and lie on a telluric absorption line.
\par
Class 2 objects correspond to ``possible'' redshifts. These include some spectra that are reasonably likely to be correct (e.g. H\&K on top of a telluric line and no other corroborating features), but also some that are little more than guesses. Class 1 is reserved for ``junk'', with no chance of obtaining a redshift. In this analysis we only consider galaxies with class 3 or 4 quality redshifts, which represent 603 of the 810 unique galaxies for which a good spectrum was extracted.
\par
We check our redshift accuracy in two ways. There are ten galaxies with redundant observations, where both observation yielded a class 3 or 4 redshift. The dispersion of the redshift differences for these objects is $\sim 150$ km/s, meaning the typical redshift uncertainty is $\sim 100$ km/s. Secondly, we deliberately re-observed 11 galaxies with redshifts available from zCOSMOS. As in \citet{2011MNRAS.412.2303B}, we find a small systematic offset, such that our redshifts are smaller than those in zCOSMOS. Over the two semesters we have 19 duplicates of this type. Omitting one outlier, the bias is $\Delta z=6.2\times10^{-4}$, corresponding to a rest-frame velocity bias of $\sim 100$ km/s at $z=0.9$. Then, we adjust the zCOSMOS redshifts by this small amount so they are consistent with ours.
\subsubsection{zCOSMOS}\label{sec-zcosmos}
\par
We also make use of the DR2 release of the 10K zCOSMOS spectroscopic survey \citep{2007ApJS..172...70L, 2009ApJS..184..218L}, from which we obtain spectra, photometry and redshifts for moderately bright galaxies ($i<22.5$) over most of the survey area. This is a sparsely-sampled redshift survey, with $\sim 40$ per cent sampling completeness. All galaxies with redshift quality greater than 2.0, and with a high probability ($>90$ per cent) of being correct (note the zCOSMOS quality flags are defined differently from our own) are used.
\par
For most of the analysis, we only consider galaxies within the field-of-view of our GMOS fields, which simplifies and homogenizes the selection function. In some cases where this is not important, however, we consider the whole catalogue. 
\begin{figure}
	\leavevmode
	\epsfysize=8cm
	\epsfbox{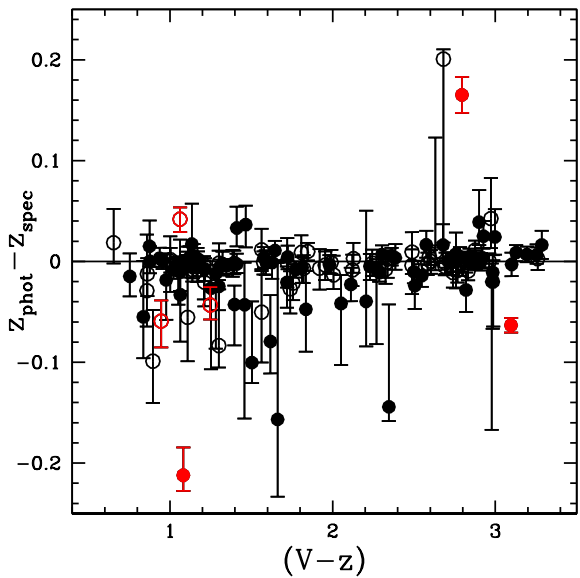} 
	\caption{For confirmed group members with $0.7<z_{\rm phot}<1.5$ we show the difference between spectroscopic and photometric redshifts, as a function of $(V-z)$ colour. Asymmetric error bars are 68 per cent confidence intervals on $z_{\rm phot}$. {\it Filled symbols} are quality class 4, which means the spectroscopic redshift is certain, while {\it open symbols} are class 3 and thus generally reliable (and used throughout this analysis). The {\it red points} identify those few group galaxies that were allocated lower priority during mask design because their $z_{\rm phot}$ is more than 2$\sigma$ away from the mean group redshift. These make a negligible contribution to the final membership.}
	\label{fig-cfz3}
\end{figure}
%
%
\section{Analysis}\label{sec-anal}
\subsection{Group Masses and Membership}\label{sec-gmem}
\par
The groups in our sample have already been robustly identified from sparse spectroscopy and deep X-ray observations. With the additional spectroscopy we do not therefore start with a group-finding algorithm, but only improve existing estimates of the centroid and the velocity dispersion. As these are confirmed, significant over-densities with extended X-ray emission and well-determined centres, the amount of possible field contamination should be substantially lower than the $\sim15$ per cent expected from a general group-finding algorithm \citep{2012ApJ...753..121K}.
\par
For each group, we start by considering all galaxies within $r=1$ Mpc and 4000 km/s (rest-frame) of the nominal X--ray group centre from \citet{2011ApJ...742..125G, 2012ApJ...757....2G}. The velocity dispersion of these galaxies is determined using the gapper estimate \citep{1990AJ....100...32B}. We also compute the (unweighted) mean spatial position of the galaxies, and the {\it rms} projected separation from this centre, which we call $R_{\rm rms}$. We then iterate this process, clipping galaxies with velocities $v>C_z\sigma$ from the median redshift, and positions $r>C_r R_{\rm rms}$ from the recomputed centre.
\par
For most groups, we use $C_z=2$ and $C_r=2$. Groups 150 and 161 are close together in redshift and position, and we adjusted these parameters to better separate them. We adopt $C_z=1.5$ and $C_r=2$ for group 150, and $C_z=1.4$ and $C_r=1.6$ for group 161. Finally, groups 121 and 213a have few members and we needed to relax $C_z=2.5$.
\par
We adopt the final $\sigma$ and $R_{\rm rms}$ from this procedure, and define group members as those within $2.5\sigma$ of the median redshift and within $r<2R_{rms}$, although the limiting radius can be varied depending on the application. Uncertainties on $\sigma$ and $R_{\rm rms}$ are then computed using a jackknife method, iterating only over this list of group members. Thus, these uncertainties do not include systematic biases due to the clipping process, which are likely to be at least $\sim 15$ per cent \citep[e.g.][]{2005MNRAS.358...71W}. 
\par 
The final velocity dispersions, reported in Table~\ref{tab-gprop}, are corrected to their rest-frame values. No explicit correction is made for the effect of redshift uncertainties on these dispersions. The dispersions of the few groups with $\sigma<200$ km/s are thus likely biased high, perhaps up to a factor of $\sim2$ \citep{2009ApJ...697.1842K}. Note that this systematic bias can be larger than the statistical uncertainties on the velocity dispersion.
\par
Group dynamical masses are computed from $\sigma$ and $R_{\rm rms}$, as in \citet{2011MNRAS.412.2303B}:
\begin{equation}
	\label{eqn-mdyn}
	M_{\rm dyn}=\frac{3}{G}R_{\rm rms}\sigma^2
\end{equation}
These masses, listed in Table~\ref{tab-gprop}, range from $\sim 2\times 10^{12}M_\odot$ to $\sim 1.8\times 10^{14}M_\odot$.
\par
Any galaxy at $0.8<z<1$, within the GMOS fields-of-view, but not assigned to a group, forms part of our comparison sample. This is not necessarily a low-density sample, as it may contain unidentified groups, but nor is it representative of the overall field, since some groups have been removed. For this paper, we will usually refer to it as the ``field'' sample for simplicity\footnote{In \citet{2011MNRAS.412.2303B} we called this the ``non-Xgroup'' sample but we decided to discard this terminology here in favour of the ``field'' sample. There is no change in its definition.}. Any difference from a truly unbiased field population is probably small, at least compared with our group sample.
\subsection{Stellar Mass Measurements and k-corrections}\label{sec-smass}
\par
We fit the spectral energy distribution for each galaxy using all available photometry, following the method described in \citet{2011MNRAS.413..996M}, but now using the updated \citet{CB07} models, commonly referred to as Charlot \& Bruzual (2007) or CB07. As previously, we assume a \citet{2003PASP..115..763C} initial mass function and, following the general procedure of \citet{2007ApJS..173..267S}, assume the star formation history of a galaxy can be represented by an exponential decay model with superposed bursts.
\par
Every model in the template suite is compared with the data, and the $\chi^2$ difference is computed. This is then used to define a weight $w_i=\exp{\left(-\chi_i^2/2\right)}$ for model $i$, and we construct a probability distribution function (PDF) for every parameter of interest by compounding these weights. We adopt the median of the PDF as our best estimate of the parameter, and the $16-84$ percentile range of the PDF is taken to represent the $\pm 1\sigma$ uncertainty. 
\par
We k-correct all colours to $z=0.9$, the redshift of interest for our survey, using the {\sc kcorrect} {\sc IDL} software of \citet{2007AJ....133..734B}, to minimize the effect of such corrections. We denote these colours as, for example, $(V-z)^{0.9}$. The $(V-z)^{0.9}$ perfectly straddles the 4000 \AA\ break at $0.8<z<1$. It is roughly comparable to rest-frame $(u-B)$, but in fact, the $V$ filter is measuring rest-frame flux intermediate between GALEX NUV and ground-based u.
\par 
The difference between the stellar masses based on the older \citet{2003MNRAS.344.1000B} models, and the new ones, are generally small \citep{2010ApJ...709..644I}. However, we have made some other, more important changes to how stellar masses are computed since our previous work in \citet{2011MNRAS.412.2303B}. First, we apply a correction to the aperture magnitudes to approximate total magnitudes, which leads to a systematic increase in stellar mass. Following \citet{2010ApJ...709..644I}, we also test the sensitivity of the $\chi^2$ distribution to small zero-point offsets, and the omission of particular filters. We find that removing the J-band data (where strong H$\alpha$ emission is often present), and applying small zero-point offsets, significantly improves the number of $\chi^2$ outliers, and thus we adopt this modification to the fits.
\par
This procedure leads to masses that are $\sim 50$ per cent larger on average from those we used in \citet{2011MNRAS.412.2303B}, with a standard deviation of about a factor $\sim 2$.
\subsection{Spectroscopic Indices}\label{sec-spec}
\par
Spectral line indices were measured for all galaxies with secure redshifts. First, the 1D spectra are shifted into each galaxy's rest-frame with the {\sc iraf} {\sc dopcor} command. The FITS files are then imported into {\sc IDL}, where the relevant spectroscopic indices can be measured using a custom-written script. The equivalent widths of the \oii and H$\delta$ lines are defined by first fitting a straight line between the centres of the red and blue continuum ranges, where the estimates for flux levels of the red and blue continuum were obtained by calculating the average flux over the wavelength ranges presented in Table~\ref{tab-cont}. Next, the flux over the line index is summed over the wavelength ranges listed in Table~\ref{tab-cont} for each spectral feature. Using the calculated linear fit to the continuum, the equivalent width for the H$\delta$ and \oii lines can be determined using the equation:
\begin{equation}\label{eqn-ew}
\mathrm{W}_\lambda = \int \left(1-\frac{\mathrm{F}_\lambda}{\mathrm{F_c}}\right) \dif \lambda,
\end{equation}
where $\mathrm{F_c}$ is the flux level of the continuum and $\mathrm{F}_\lambda$ is the total flux per unit wavelength.
\par
For this paper, we adopt the sign convention such that a positive \oii equivalent width indicates an emission feature, while a positive \hdelta equivalent width corresponds to an absorption feature.
\begin{table}
	\begin{tabular}{|l|l|l|l|}
 	\hline
 	Index & Blue Continuum & Line Definition & Red Continuum \\
		\hline 
		\oii & 3653 - 3713 & 3722 - 3733 & 3741 - 3801 \\
		H$\delta$ & 4030 - 4082 & 4082 - 4122 & 4122 - 4170 \\
 	\hline
 	\end{tabular}
	\caption{The table list the wavelength intervals used for the estimation of the blue and red continuum, as well as the line definition for the spectroscopic indices used in this paper. The values in the table are provided in units of Angstroms.}
	\label{tab-cont}
\end{table}
\par
\subsubsection{Weighted-Median and Stacked-Spectra Methods}\label{sec-twomethod}
\par
Since the signal to noise ratio for these faint spectra are relatively low ($S/N\sim 1$ per pixel), any measurements of spectroscopic indices (including their derived quantities, like the star formation rate) can have considerable uncertainty. It is therefore valuable to consider quantities averaged over various subpopulations, and we do this in two different ways in this paper.
\par
For the first method, the relevant indices for each galaxy spectrum are measured individually, using the procedure outlined in \S~\ref{sec-oii}. We then compute a median that includes the galaxies' spectroscopic and mass-completeness weights (the derivation of these weights will be further described in \S~\ref{sec-zsuc}). For this weighted-median method, the relevant quantities (such as \oii luminosity) are first sorted in order of increasing value. Then, the total weight is calculated. The point at which the cumulative weighting is equal to half of the total weight is considered to be the weighted-median value. The uncertainties in this method can be determined with the bootstrap method. The distribution is randomly re-sampled 500 times and the median value re-calculated for each iteration. The standard deviation of this distribution of median values can provide a robust error estimate for the original measurement.
\par 
The second method is to stack the individual spectra and measure the feature on the combined spectrum. First, the spectra are multiplied by their total weights, which is a combination of their spectroscopic and mass-completeness weighting. Then, the spectra are combined by taking the average flux of each pixel, using the {\sc iraf} function {\sc scombine}. The resulting stacked spectra is used for the subsequent calculations for the properties of this bin. The measurement errors in the equivalent widths of the stacked spectra can be calculated using the formula from \citet{2006AN....327..862V}.
\begin{equation}
\sigma(\mathrm{W}_\lambda) = \sqrt{1+\frac{\overline{\mathrm{F_c}}}{{\overline{\mathrm{F}}}}}\ \frac{\Delta\lambda-\mathrm{W}_\lambda}{\mathrm{S/N}}
\end{equation}
where the S/N term is the total signal to noise ratio over the line index in question. This is defined as the mean value of the flux per pixel, divided by the variance and multiplied by the square root of the number of pixels in the line index definition. Note that this estimate represents only the uncertainty in the mean value, and does not include the variance associated with the distribution of pixel values entering the stacking process.
\subsubsection{Active Galactic Nuclei}
\par
We identify strong AGN by cross correlating with the point source catalogue from deep {\it Chandra} observations within 5\arcsec. Only seven galaxies have such a match, and they are identified in all appropriate figures as symbols with a thick, black square around them. Weak, narrow-line AGN, or heavily-obscured AGN could still be present in the sample. Radio-loud AGN could also be present, but these are not expected to bias SFR measurements.
\subsection{Completeness Corrections}\label{sec-weights}
\subsubsection{Spectroscopic and Redshift Completeness}\label{sec-zsuc}
\par
There are several potential sources of incompleteness, which we will consider in turn.
\par
First is the sampling completeness, meaning the fraction of all likely group members in the photometric sample for which we obtained a spectrum. This is primarily dependent on magnitude and position, with no significant colour dependence. For the group sample, we consider all galaxies that have a photometric redshift consistent with the final group redshift, within twice the 68 per cent confidence limits of the photometric redshift. The targeting completeness within $R_{\rm rms}$ is computed on an individual group basis, in three $r$ magnitude bins (22--23.25, 23.25--24, 24--24.75). Note that our approach here is different from that described in \citet{2011MNRAS.412.2303B}. For the field sample, we simply measure the fraction of galaxies with photometric redshifts between $0.8$ and $1.0$ and found within our GMOS fields of view, with a spectrum. A linear function of this fraction and $r$ magnitude is fitted.
\par
A second source of incompleteness is the failure to obtain a reliable redshift from a spectrum. As in \citet{2011MNRAS.412.2303B}, our success rate is very high, $>80$ per cent for $i<24$. There is no significant colour dependence of this fraction. Many of the failures are likely $z>1.4$ galaxies (where \oii is redshifted out of the observed wavelength range) and thus outside the redshift range of interest here, so we apply no correction for this.
\par
Although the photometric redshift pre-selection greatly improves the efficiency of our observations, it introduces a potential bias. In particular, the process is sensitive to catastrophic failures, where objects in which the photometric redshift differs from the true redshift by many times the estimated uncertainty.  The contribution of these outliers will not be appropriately recovered with our completeness weighting.  To address this, in Figure~\ref{fig-cfz3} we show the difference between the spectroscopic and photometric redshift for all confirmed group members (see \S~\ref{sec-gmem}) with secure spectroscopic redshifts and $0.7<z_{\rm phot}<1.5$, as a function of their $(V-z)$ colour. Red points identify ``priority 2'' galaxies, which are those lower priority targets with $z_{\rm phot}$ more than 2$\sigma$ away from the group redshift, but still within $0.7<z_{\rm phot}<1.5$. These seven galaxies contribute $\sim 4.0\pm1.5$ per cent to the final group sample. From the present, small sample, these galaxies have colours representative of the rest of the group population. Moreover, only two of them (representing 1.2 per cent of the group sample) have stellar masses greater than the limits to which we consider our results complete (see \S~\ref{sec-mcomp}). Generally, however, we sample the priority $=2$ population more sparsely than our primary sample. The sampling completeness of priority $=2$ galaxies is $\sim 0.15$, or about $0.2$ times that typical of the top priority population, with a range of $10$--$30$ per cent, depending on the group. Thus, we might expect that even after completeness corrections described above have been applied, we may be missing $\sim 6$ per cent of the group population above our completeness limit, due to this bias.
\par
Finally, when computing volume-normalized quantities like the stellar mass function for the field sample, we include a weight to account for the fact that the lowest luminosity objects are drawn from a smaller volume.
\subsubsection{Mass Completeness}\label{sec-mcomp}
\begin{figure}
	\leavevmode
	\epsfysize=8cm
	\epsfbox{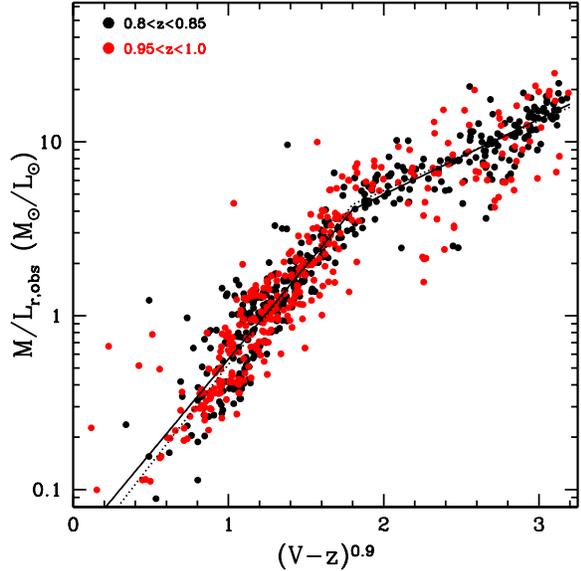} 
	\caption{We show the stellar mass-to-light ratio as a function of colour, for galaxies with secure redshifts in our sample. The mass is derived from SED fitting using \citet{CB07} models with a \citet{2003PASP..115..763C} IMF, and the luminosity here does not include a k-correction. Black points represent galaxies at the lower end of our redshift range, $0.8<z<0.85$, while the red points correspond to the upper end at $0.95<z<1.0$. The solid and dotted lines (barely distinguishable over most of the range) are fits to the median $M/L_r$ for the low and high redshift samples, respectively. The fit is bilinear, with a different slope for red and blue galaxies.} 
	\label{fig-mlr_colourfig}
\end{figure}
\begin{figure}
	\leavevmode
	\epsfysize=8cm
	\epsfbox{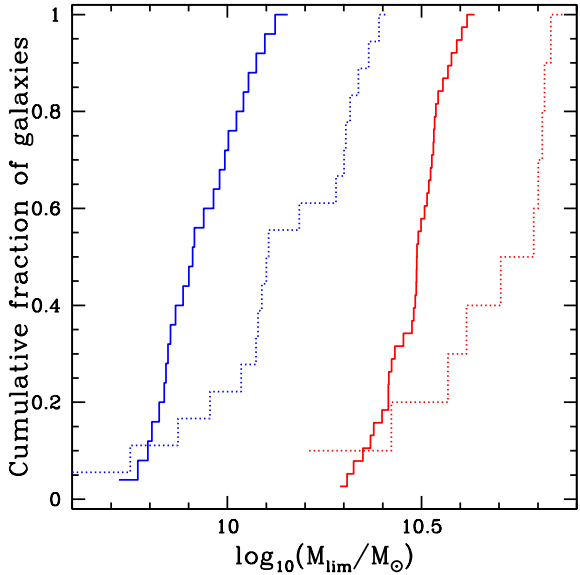} 
\caption{Each curve shows the cumulative distribution of limiting stellar mass (i.e. at a given stellar mass, the fraction of the $M/L_r$ distribution that will be included in the sample) for a population of fixed $(V-z)^{0.9}$ colour and redshift. This is derived from the luminosity limit and the distribution of $M/L_r$ (see Fig~\ref{fig-mlr_colourfig}) for that population. Weighting galaxies of that mass by the inverse of this number then accounts statistically for galaxies with high $M/L_r$ ratios that may be absent from the sample. The blue curves represent blue galaxies $(V-z)^{0.9}\sim 1.7$, while the red lines correspond to $(V-z)^{0.9}\sim 2.9$. Solid and dashed lines represent the lower and upper redshift range of the data, respectively ($z=0.8$ and $z=0.95$).}
\label{fig-calc_cweightfig}
\end{figure} 
\par
The $r$-selection of our survey introduces strong colour-dependence in the stellar mass completeness. In Figure~\ref{fig-mlr_colourfig} we show the stellar mass-to-light ratio as a function of $(V-z)^{0.9}$ colour, where the luminosity $L_r$ is computed from the selection magnitude ($r$) and the luminosity distance, without any k-correction. Therefore, the $M/L_r$ ratio directly relates the selection magnitude to the physical quantity of interest here (stellar mass), with distance as the only other relevant parameter. We divide the galaxies into those at the lower end of our redshift range ($0.8<z<0.85$, black points) and at the upper end ($0.95<z<1.0$, red points).
\par
The strong correlation with colour primarily reflects the different star formation histories. At a fixed colour, there is a range of $M/L_r$ ratios that reflects a variety of effects, including physical effects like dust and metallicity variations. There are also variations in the rest-wavelength coverage, which depends on redshift and  SED shape. Finally, uncertainties in the SED-fitting procedure, resulting from photometric uncertainties, or the random sampling of bursts in the population modeling for example, will also contribute. The lines show the median $M/L_r$ at the lower (solid line) and upper (dotted line) ends of the redshift range. There is no significant difference between the two. The luminosity limit of our sample is well defined, by the $r<24.75$ sample selection. This limit then corresponds to a range of stellar masses that depends strongly on colour. At fixed $(V-z)^{0.9}$ colour, the maximum value of $M/L_r$ determines the 100 per cent mass completeness limit. For example, the reddest galaxies have $M/L_r$ as high as $30$, and thus the sample is only fully complete for $M_{\rm star}>10^{10.8}M_\odot$. 
\par
However, this is unnecessarily conservative. With the assumption that the $M/L_r$ distribution, at fixed $(V-z)^{0.9}$ colour, is not a strong function of stellar mass, we can use the distribution shown in Figure~\ref{fig-mlr_colourfig} to statistically correct for the missing, high $M/L_r$ galaxies.
We construct the cumulative distribution of limiting masses, given by the $r$-band luminosity limit at a given redshift multiplied by the $M/L_r$, in bins of colour and redshift. An example is shown in Figure~\ref{fig-calc_cweightfig}. Each curve shows, at a given mass, what fraction of the $M/L_r$ distribution (for galaxies of that type and redshift), will be included in the sample. Weighting galaxies of that mass by the inverse of this number then accounts statistically for galaxies with high $M/L_r$ ratios that may be absent from the sample. If we are willing to accept statistical corrections as large as a factor $\sim 5$, we can extend the mass limit of blue galaxies by a factor $\sim 4$, and of red galaxies by a factor $\sim 2$.
\par
This correction is not strictly a statistical weight, but a systematic correction. It means we are assuming that the distribution of dust, metallicity, and SED-fitting uncertainties is mass-independent for galaxies of a given colour over a relatively small mass range. This does mean that the galaxies we are missing are preferentially those with high $M/L_r$ ratios, e.g. those with more dust than average. These may not be representative of the whole galaxy population at that mass and will remain a caveat to our methodology.
\subsection{Galaxy Sub-Populations}\label{sec-passive}
\par
It is now well known that most galaxies tend to populate either the ``passive-sequence'' of galaxies with negligible present star formation, or the ``blue cloud'' of star-forming galaxies, with star formation rates $SFR\propto M_{\rm star}^\alpha$, where $\alpha$ is generally negative but not too far from unity \citep[e.g.][]{2007ApJ...660L..43N, 2012ApJ...754L..29W}. It is of interest to consider our sample in terms of these two populations and any galaxies that might be intermediate between the two.
\par
\begin{figure}
	\leavevmode
	\epsfysize=8cm
	\epsfbox{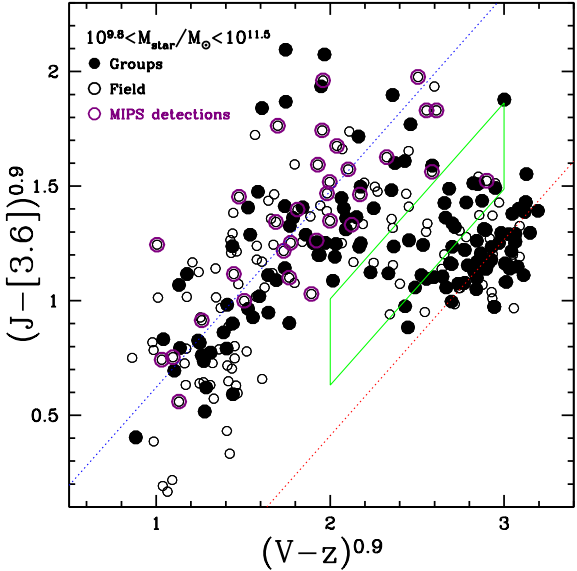} 
	\caption{Galaxies at $0.8<z<1$ are shown in the colour-colour plane, aimed at separating dust-reddened spiral galaxies from the truly passive population \citep[e.g.][]{2009MNRAS.398..754B, 2011MNRAS.412.2303B}. Confirmed group members are shown as {\it filled symbols}, while field members within our GMOS fields-of-view are shown with {\it open symbols}. The fits to the red and blue populations, are shown in the dotted lines. The thick green lines represent the boundaries defining the intermediate, green population. Galaxies with 24 micron MIPS detection within 3$''$ are outlined in purple.}
	\label{fig-ccbest}
\end{figure}

While this choice could be made based directly on estimates of the SFR, there are advantages to considering a classification based on more directly observable quantities. It has been shown by us \citep{2009MNRAS.398..754B, 2011MNRAS.412.2303B} and others \citep[e.g.][]{2005ApJ...624L..81L, 2005A&A...443..435W, 2012ApJ...745..179W} that the active and passive populations are well-separated in a colour-colour plane, as shown here in Figure~\ref{fig-ccbest}. The x-axis colour, $(V-z)^{0.9}$, brackets the 4000 \AA\ break and is a good indicator of luminosity-weighted age, while the y-axis colour $(J-[3.6])^{0.9}$, is sensitive to dust extinction. We plot all galaxies with good redshifts $0.8<z<1$ within a broad mass range $M_{\rm star}>10^{9.8}M_\odot$. The figure includes all group and field galaxies within the GMOS fields-of-view. We highlight with purple circles galaxies detected in deep MIPS data, with 24$\mu$m flux greater than $0.071$ mJy \citep{COSMOS_Spitzer}.
\par
\begin{figure}
	\leavevmode
	\epsfysize=8cm
	\epsfbox{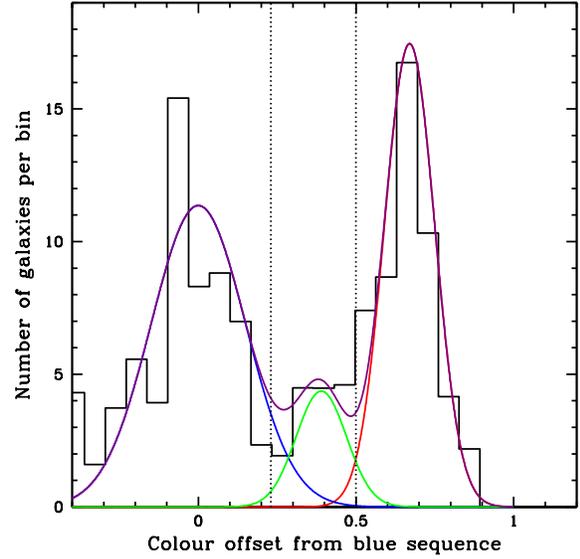} 
	\caption{The distribution of group galaxies positions (with masses $9.8<\log(M_{\rm star}/M_\odot)<11.5$) in the colour-colour plane, measured as the perpendicular offset from the peak of the blue sequence.  The histogram data includes the relevant statistical weights. Also plotted are Gaussian fits to the red, blue, and green population, and the summed function of all three fits. The vertical lines represent our identification of the intermediate, green population.}
	\label{fig-ccproj}
\end{figure}

Figure~\ref{fig-ccbest} demonstrates the well-known fact that selecting passive galaxies based only on an optical colour straddling the 4000 \AA\ break (e.g. $(V-z)^{0.9}>2.4$) would include a significant population of dust-reddened galaxies with significant star formation. Therefore, using optical data alone in the colour-colour plane, we seek to identify truly passive galaxies with no contamination from very dusty galaxies. First, a line is fit to the main distribution of galaxies, shown in Figure~\ref{fig-ccbest} as the blue dotted line\footnote{The equation of this line is $y=0.89x-0.28$, where $x=(V-z)^{0.9}$ and $y=(J-[3.6])^{0.9}$.}. The red dotted line in Figure~\ref{fig-ccbest} has an identical slope, but with an offset of $-1.02$ mag to fit the second concentration of galaxies observed. Next, the distance of each galaxy from the two main distributions is measured as the perpendicular distance from these lines. The distribution of these distances is shown in Figure~\ref{fig-ccproj}, with a clear separation between the two peaks evident. We also observe an intermediate population, located in between the main distributions. We will refer to these as the ``green'' galaxies, as in \citet{2011MNRAS.412.2303B}. After fitting the three populations with Gaussian functions, we define the boundaries of the green population so it has the minimal overlap with the passive and star-forming sequence. This is shown by the dotted lines in Figure~\ref{fig-ccproj}. The resulting green box in Figure~\ref{fig-ccbest} represents our definition of the ``transition'' population. Note that there are additional vertical limits drawn using in the $(V-z)^{0.9}$ axis, in order to prevent very blue or very red objects from being identified as green galaxies. Similar attempts to identify green galaxies have been made recently by \citet{2011MNRAS.411..929G} and \citet{2012ApJ...745..179W}, though the definitions differ from ours.
\par
\begin{figure*}
	\leavevmode
	\epsfysize=8cm
	\epsfbox{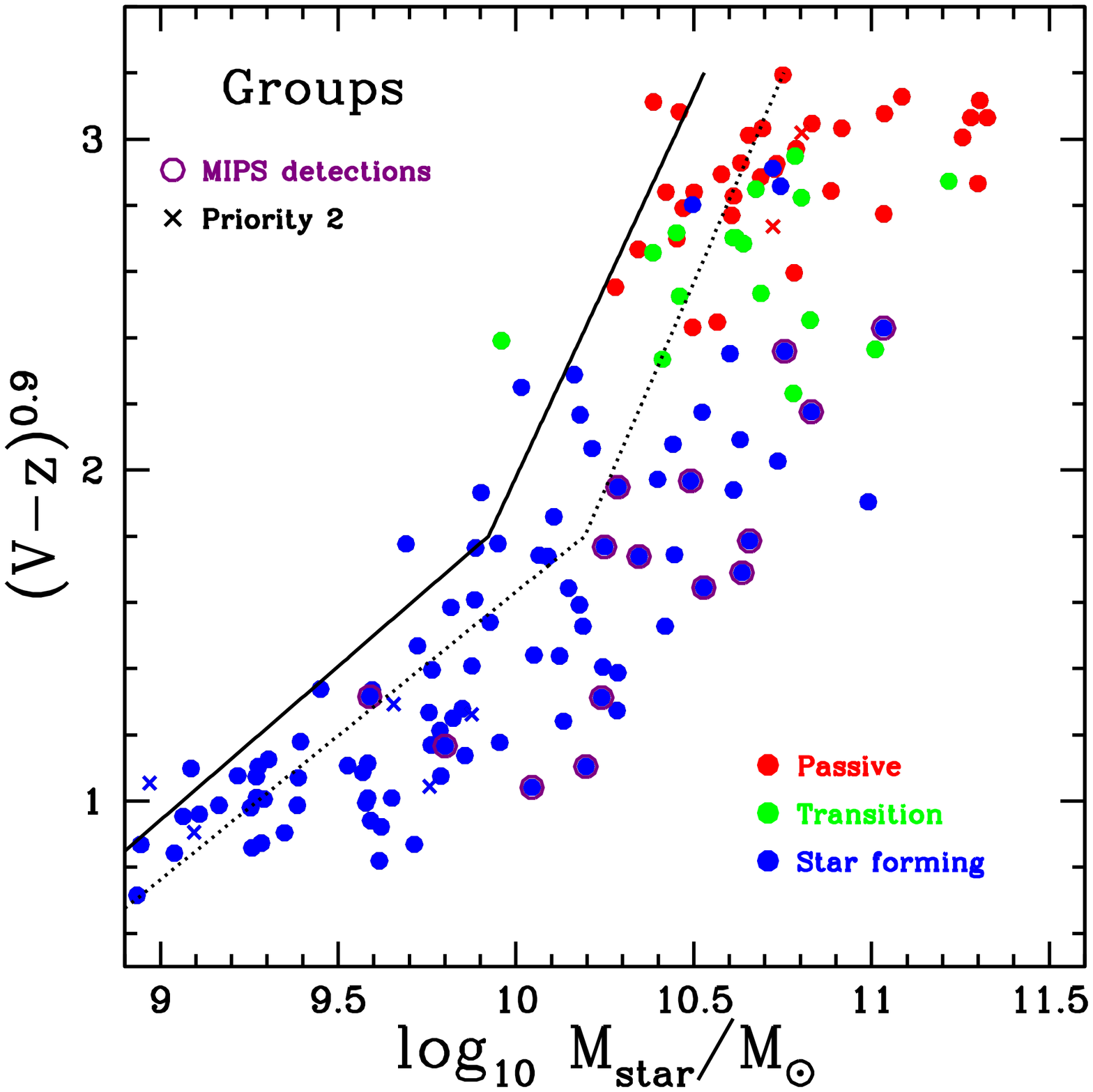}
	\epsfysize=8cm
	\epsfbox{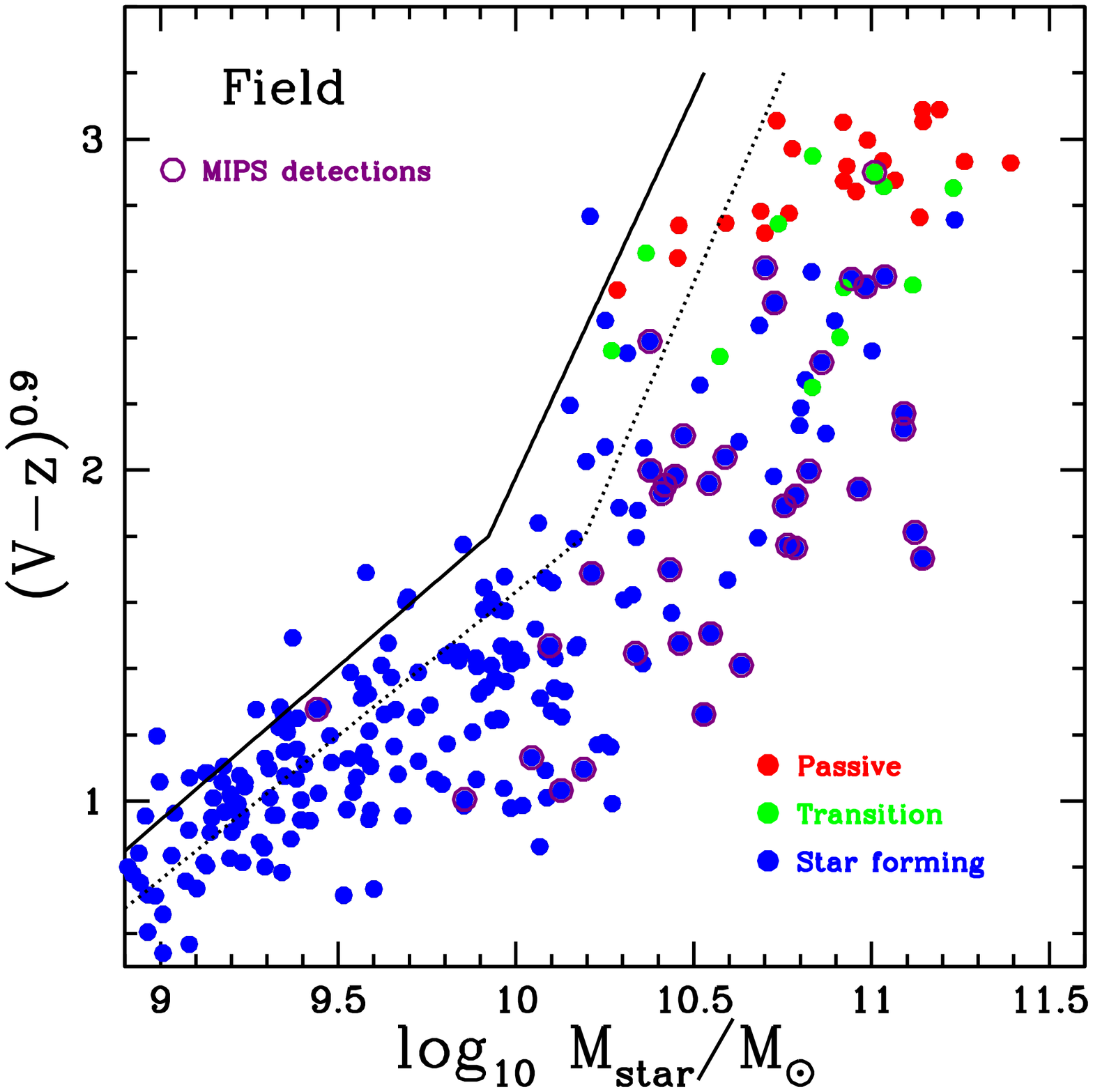} 
	\caption{The $(V-z)^{0.9}$ colour is shown as a function of stellar mass, for group galaxies (left panel) and for the field popluations (right panel). Galaxies are divided into three types, based on their location in colour-colour space (Figure~\ref{fig-ccbest}). Red, green and blue galaxies are colour-coded appropriately. The solid line and dotted line represent the stellar mass limit represented by the median $M/L_r$ ratio and our $r=24.75$ magnitude selection, at $z=0.8$ and $z=1$, respectively. Galaxies with 24 micron MIPS detection within 3$''$ are outlined in purple. The few {\it cross symbols} in the left panel correspond to group galaxies from our more sparsely sampled, priority 2 population, in which the $z_{\rm phot}$ is more than 2$\sigma$ away from the group redshift. Most of these are low-mass (faint) galaxies, where even the primary sample is highly incomplete, except for the bluest galaxies.}
	\label{fig-cmass_z}
\end{figure*}

In Figure~\ref{fig-cmass_z}, we show the $(V-z)^{0.9}$ colour dependence on stellar mass. Red, green and blue galaxies as defined above are represented with like-coloured symbols. The straight lines represent the adopted stellar mass limit, defined as the mass corresponding to our $r=24.75$ mag selection limit at the $M/L_r$ ratio for which 50 per cent of our galaxies at that colour are expected to have lower $M/L_r$. The solid and dotted lines represent the limits at $z=0.8$ and $z=1$, respectively. We also show, as crosses, the seven group galaxies that originated from the priority $=2$ sample. Most of these are low-mass galaxies, below the limit where we are complete except for the very bluest systems. 
\begin{table*}
\begin{minipage}{125 mm}
	\begin{tabular}{|l|l|l|l|l|l|l|l|l|}
 	\hline
 	Log Stellar Mass & Blue Group & Blue Field & Green Group & Green Field & Red Group & Red Field\\
		\hline 
		9.5 - 9.8 & 21 & 26 & - & - & - & - \\
		9.8 - 10.1 & 17 & 50 & - & - & - & - \\
		10.1 - 10.4 & 20 & 28 & 2 & 2 & - & - \\
		10.4 - 10.7 & 12 & 16 & 7 & 1 & 14 & 4 \\
		10.7 - 11.0 & 13 & 15 & 4 & 5 & 11 & 6 \\
		11.0 - 11.3 & 1 & 5 & 2 & 2 & 11 & 7 \\
		11.3 - 11.6 & - & - & - & - & 1 & 1 \\
 	\hline
 	\end{tabular}
	\caption{The number of galaxies in each mass bin, based on the colour definitions in Figure~\ref{fig-ccbest}. The count is restricted to galaxies in the GMOS fields-of-view and with total weights (spectroscopic and mass completeness) of between 0 and 5. Within the GMOS field of view, for 73 of the 304 galaxies in the sample, the zCOSMOS spectra were used.}
	\label{tab-mass}
\end{minipage}
\end{table*}
\subsection{Star Formation Rates}\label{sec-oii}
\par
While SFR can be estimated from our SED-fits, these can be difficult to interpret, especially for low-SFR galaxies which are undetected in the rest-UV.  
The SFR can instead be obtained directly from the \oii feature, as its luminosity is coupled to the H II regions, the sites of star-formation inside galaxies. However, this line is more strongly affected by dust extinction and metallicity than comparable indicators, like the H$\alpha$ recombination feature. To compensate, a recent paper by \citet{2010MNRAS.405.2594G} presents an empirical conversion between observed \oii luminosities and SFR at $z=0.1$ in a statistical manner, calibrating the result to values derived using H$\alpha$ and UV data:
\begin{equation}\label{eqn-dgg}
\frac{\mathrm{SFR}}{\mathrm{M}_\odot\ yr^{-1}} = \frac{\mathrm{L_{[O\ II]}}}{3.80\times10^{40}\mathrm{erg\ s^{-1}}}\frac{1}{a\ \mathrm{tanh}[(x-b)/c] + d},
\end{equation}
with $a=-1.424,\ b=9.827,\ c=0.572,\ d=1.700,\ x=\log (M_\mathrm{star}/M_\odot)$.
\par
We can measure the \oii luminosity by multiplying the equivalent width measurement with the luminosity of the continuum at its redshifted position. In the range $0.8<z<1.0$, the $r$ and $i$ bands bracket the \oii line and the continuum luminosity is estimated by interpolating between the total magnitudes in those bands, measured from the broad-band images. As long as the locally-calibrated relation in Equation~\ref{eqn-dgg} holds at $z=1$, these SFR are dust-corrected. However, this only includes the average dust correction for galaxies of that stellar mass and does not include any individual correction based on the observed properties.
\par
A comparison of the \oii - derived SFR using this method and the FUV + IR SFR for our sample show a systematic normalization offset between the two, as noted by others \citep[e.g.][]{2011ApJ...735...53P, 2011ApJ...730...61K}.  It is not clear which method gives the right answer, but in principle the FUV + IR should be the most direct, with the IR bolometric correction being the largest systematic uncertainty. At this redshift range, $0.8<z<1.0$, the IR luminosity estimated from the 24 micron line should remain accurate \citep{2011A&A...534A..15M}.
\par
The \citet{2010MNRAS.405.2594G} calibration would be expected to underestimate the true SFR if these relatively massive galaxies at $z=1$ are dustier than galaxies of similar mass at $z=0$, which may not be unreasonable. Recent studies have found that the average dust attenuation may peak at $z\sim1$ \citep{2012A&A...539A..31C}. Thus, we elect to multiply the \oii - derived SFR by a factor of 3.1 in all subsequent analysis, as explained in Appendix~\ref{app1}. The difference between these two methods will be explored in a subsequent paper, Mok et al. (in prep), which will also compare the results from other star formation indicators on our sample, including the use of SED-fitting. 
%
%
\begin{figure}
	\leavevmode
 	\epsfysize=8cm
	\epsfbox{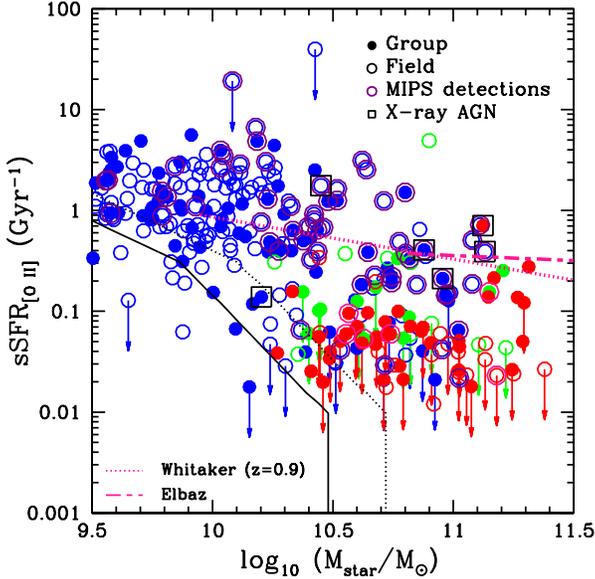}
	\caption{The specific star formation rate measured from the \oii feature is shown as a function of stellar mass for galaxies at $0.8<z<1$ with secure redshifts. Group members are shown as {\it filled symbols}, and field members within our GMOS fields-of-view are shown as {\it open symbols}. Only galaxies with total weight (spectroscopic and mass-completeness) of between 1 and 5 are included. The solid line and dotted lines represent our 50 per cent completeness limit due to the $M/L_r$ distribution at fixed sSFR, at $z=0.8$ and $z=1$, respectively. Points are colour-coded according to their classification as red, blue, and green. Galaxies with significant X-ray sources within 5$''$ are outlined in black, while galaxies with 24 micron MIPS detection within 3$''$ are outlined in purple. The star-forming sequence from \citet{2007A&A...468...33E} {\it (thick pink dashed line)} and \citet{2012ApJ...745..179W} {\it (thin pink dashed line)} at $z=0.9$ are included for comparison, up to their stated mass completeness limits. Note that those galaxies with a negative \oii equivalent width or with \oii detections less than the measurement error are converted into upper limits.}
	\label{fig-ssfr_mass}
\end{figure}
\begin{figure}
	\leavevmode
	\epsfysize=8cm
	\epsfbox{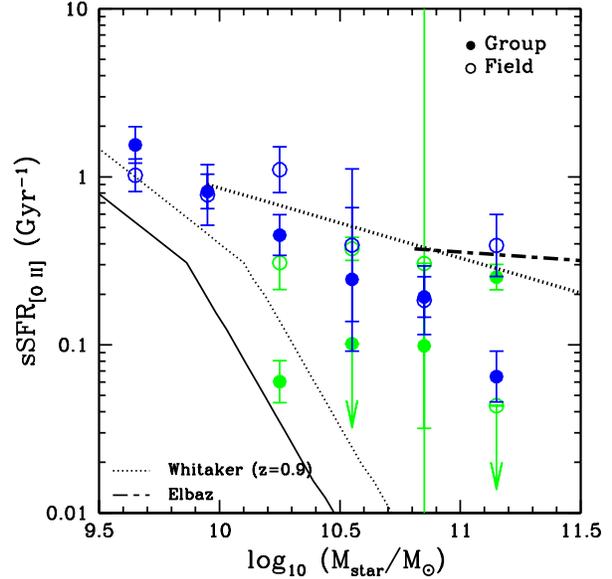}
	\caption{The binned median sSFR is shown as a function of stellar mass for galaxies at $0.8<z<1$, within the GMOS fields-of-view and total weights (spectroscopic and mass-completeness) between 1 and 5. Group galaxy bins are shown with {\it filled symbols} and field galaxies as {\it open symbols}. The solid line and dotted lines represent our 50 per cent completeness limit due to the $M/L_r$ distribution at fixed sSFR, at $z=0.8$ and $z=1$, respectively. The star-forming sequence from \citet{2007A&A...468...33E} {\it (thick dashed line)} and \citet{2012ApJ...745..179W} {\it (thin dashed line)} at $z=0.9$ are included for comparison, up to their stated mass completeness limits. The blue points are the binned values for the blue galaxies and the green points indicate green galaxy bins. For the green galaxies, cases where the weighted median would indicate an undetected \oii emission are shown as upper limits. Note that for the most massive blue group bin, which contains only one galaxy, the error of the measurement is shown.}
	\label{BG05_sSFR}
\end{figure}
\section{Results}
\subsection{The Correlation between sSFR and Stellar Mass}\label{sec-oii_ssfr_m}
\par
In Figure~\ref{fig-ssfr_mass} we show the specific star formation rate as a function of stellar mass, considering galaxies within the GMOS fields of view and with secure redshifts $0.8<z<1$. Galaxies are colour-coded according to their classification into red, green, and blue, as described in \S~\ref{sec-passive}. The solid and dotted lines represent the completeness limits at $z=0.8$ and $z=1$, respectively.
\par
We observe that the blue galaxies form a well-defined sequence, in good agreement with previous results from \citet{2007A&A...468...33E} at the high mass end and \citet{2012ApJ...745..179W} above their completeness limit of $M_{\rm star}\sim10^{10}M_\odot$. Our data extend this trend to lower masses, $M_{\rm star}\approx 10^{9.5}M_\odot$. In addition, we observe that the green galaxies are not well separated from the red population, as the \oii feature is rarely detected in either population. A small number of blue galaxies also do not have a detected \oii emission. At the high mass end, this may be caused by the \oii luminosity being dominated by the continuum component and difficulties in the detection of relatively weaker \oii feature from the spectra.  Variations in dust extinction from the average, and the large uncertainties especially for fainter galaxies, will also contribute to variations from a simple correlation between colour-class and line strength.
\par
The slope of the sSFR-stellar mass relationship potentially holds important clues about the processes regulating galaxy formation, and can put very strong constraints on feedback models \citep[e.g][]{2012MNRAS.422.2816B}. However, the measured slope will depend sensitively on the completeness limits and the definition of star-forming galaxies. In particular we are certainly missing galaxies with substantial SFR at $M_{\rm star}<10^{10}M_\odot$, which would mean the overall slope of the star-forming sequence is likely to be flatter than observed here, despite the inclusion of statistical weights to partially offset this. 
\par
In Figure~\ref{BG05_sSFR}, we show the weighted median sSFR for blue and green galaxies, stacked in equal mass bins, with errors calculated using the bootstrap method. For the most massive blue group bin, which contains only one galaxy, the error of the measurement is shown. For the green galaxies, cases where the weighted median would indicate an undetected \oii emission are shown as upper limits. We observed that for $M_{\rm star}>10^{10.3}M_\odot$, the blue and green field populations show slightly higher sSFR than their group counterparts. Below $M_{\rm star}\sim10^{10.3}M_\odot$, the blue group and field population do not show a statistically significant difference, which may be caused by incompleteness in our sample, since we could only detect galaxies with high sSFRs in this mass range.
\par
\begin{figure}
	\leavevmode
	\epsfysize=8cm
	\epsfbox{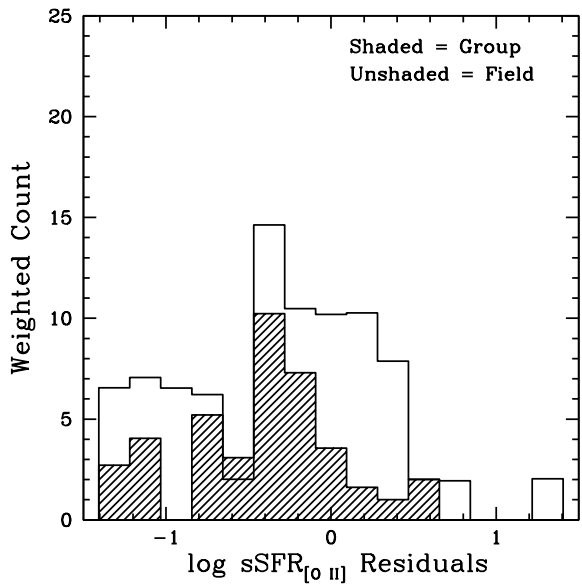}
	\caption{A histogram showing the residuals to the \citet{2012ApJ...745..179W} line at $z=0.9$, shown in Figure~\ref{BG05_sSFR}, for the high mass blue galaxy population ($M_{\rm star}>10^{10.3}M_\odot$). The data includes the spectroscopic and mass-completeness weights, drawn from the same population as Figure~\ref{fig-ssfr_mass}. The histogram for the group galaxies is shaded, while the field galaxies are unshaded.}
	\label{Histo_Res_LM}
\end{figure}

To determine the cause for this possible divergence in the sSFR of group and field population, we used the \citet{2012ApJ...745..179W} relationship between stellar mass and sSFR at $z=0.9$ and calculate the difference between their sequence and the individual galaxies' sSFR. The distribution of the residuals for the high mass blue galaxy population ($M_{\rm star}>10^{10.3}M_\odot$) is shown in Figure~\ref{Histo_Res_LM}, including the relevant spectroscopic and mass-completeness weights. Qualitatively, the main difference between the two shapes is the lack of high sSFR galaxies in the group sample. Quantitatively, the two distributions can be compared by calculating their weighted means, with the associated error provided using the bootstrap method. The field population has a mean log residual of $0.27\pm0.1$, while the mean residual for the group population is $-1.48\pm1.6$. Performing the standard Kolmogorov-Smirnov (KS) test on the unweighted sample shows that the group and the field have a 17\% chance of originating from the same distribution. We conclude, therefore, that the small offset we observe between the star-forming main sequence of $z\sim1$ group and field galaxies is not statistically significant.
\subsubsection{Stellar Mass Functions}
\par
\begin{figure}
	\leavevmode
	\epsfysize=8cm
	\epsfbox{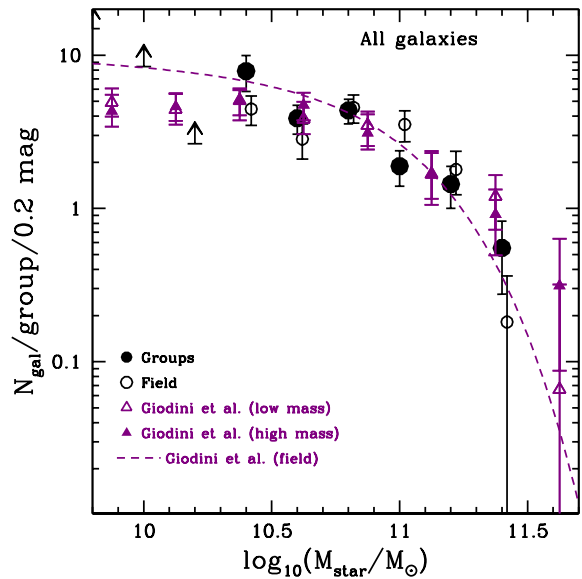} 
	\caption{The stellar mass function for all group galaxies in our sample is shown as the solid black circles, and with lower limits indicating where our sample is incomplete for red galaxies. {\it Open symbols} represent the field sample within the GMOS fields of view, slightly offset horizontally for clarity. The {\it filled open purple triangles} represent massive ($M_{\rm star}>6\times10^{13}M_\odot$) and low-mass groups from the photometric analysis of \citet{2012A&A...538A.104G}, while the {\it dashed line} represents their field sample. The Giodini data are all arbitrarily renormalized to match our data at $M_{\rm star}\sim10^{10.8}M_\odot$.}
	\label{fig-mass_f_total}
\end{figure} 
In this section we consider the stellar mass function of all galaxies in our spectroscopic group and field samples. We will compare extensively with the results of \citet{2012A&A...538A.104G}, who published mass functions for a similar (and partially overlapping) sample of groups, based on photometric redshifts only. Benefiting from their larger sample, they were able to further divide the group sample into low mass and high mass subsamples (divided at $M_{\rm star}=6\times10^{13}M_\odot$). As groups in our sample come from both sides of this division, we compare our data to both results. In addition to the deep spectroscopy, the novelty of our present work is the colour division of the galaxy population into red, blue, and green. By contrast, Giodini et al. had only divided their sample into two populations based on SED-type: passive and star forming.
\par
We begin by considering the stellar mass function of all group galaxies, computed by counting the number of galaxies in a given stellar mass bin, and applying the weights discussed in \S~\ref{sec-weights} to correct for sampling and $M/L_r$ incompleteness. We normalize each bin by the number of groups in our sample that potentially contribute (depending on their redshift), so the mass function shows the average number of galaxies per group. Then, we restrict the mass range so that all bins have contribution from at least four groups.
\par
In general, only the lowest-mass bin includes contributions from fewer than half our groups. The uncertainties in this and all subsequent mass functions are dominated by the uncertainties in the stellar masses themselves, which are systematic in nature. We account for these by simply comparing the mass function computed with all stellar masses at their $1\sigma$ upper and lower values.
\par
The resulting mass function is shown in Figure~\ref{fig-mass_f_total}, as the solid black circles with error bars. The two lower limits correspond to measurements in mass bins for which our sample is incomplete for red galaxies, even after weighting. The open circles show the corresponding mass function for the field population, considering only the sample within the GMOS fields of view. These are arbitrarily renormalized to match our ``per-group'' normalizations at $M_{\rm star}\sim10^{10.8}M_\odot$.
The filled and open purple triangles represent massive and low-mass groups from \citet{2012A&A...538A.104G}, while the dashed line represents their field sample. Again, all are renormalized to match at $M\sim10^{10.8}M_\odot$. We will retain this normalization in subsequent plots, where the galaxy population is divided by type. The mass function shapes are generally in excellent agreement over the full mass range, though the mass function of our field sample appears to be somewhat flatter than that of \citet{2012A&A...538A.104G}. 
\begin{figure}
	\leavevmode
	\epsfysize=8cm
	\epsfbox{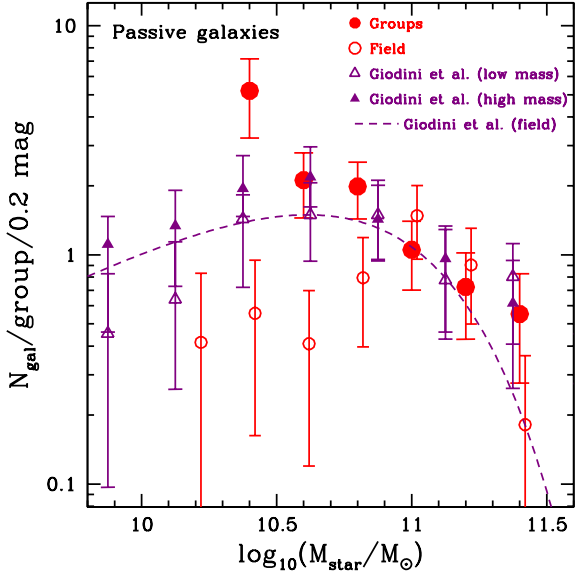} 
	\caption{The stellar mass function for passive group galaxies in our sample is shown as the filled red points; the field sample is shown with open circles. This is again compared with the high- and low-mass photometric group samples \citet{2012A&A...538A.104G}, shown as the filled and open triangles, respectively, as well as their field ({\it dashed line}). The normalization of the Giodini et al. data are all fixed according to the {\it total} mass functions. We find a steep stellar mass function for passive galaxies, with no indication for a turnover in the group population.}
	\label{fig-mass_f_red}
\end{figure}
\subsubsection{Type-Dependent Mass Functions}
\begin{figure}
	\leavevmode
	\epsfysize=8cm
	\epsfbox{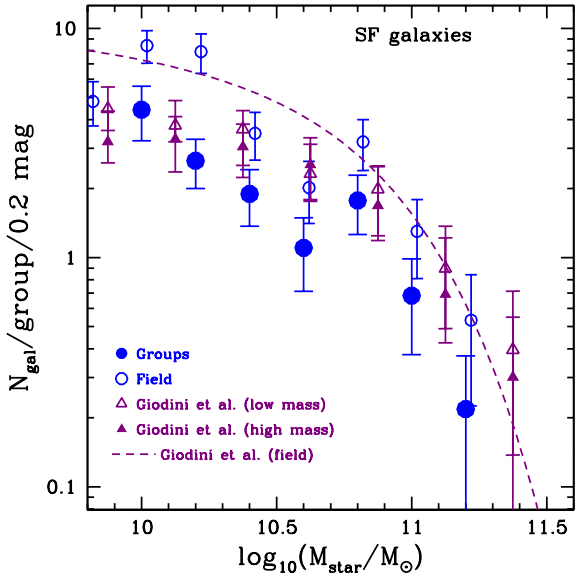} 
	\caption{As Figure~\ref{fig-mass_f_red}, but for the star-forming galaxy population, shown in blue. Purple triangles represent the groups from \citet{2012A&A...538A.104G} as in previous figures; note that their population of star-forming galaxies includes our green galaxies, which may explain the discrepancy at higher masses. The Giodini et al. field mass function for blue galaxies is shown as the dashed line.}
	\label{fig-mass_f_blue}
\end{figure} 
\par
We now divide the stellar mass functions by galaxy type as defined in \S~\ref{sec-passive}, and start with the red population in Figure~\ref{fig-mass_f_red}. We only show data to the limit where our completeness drops below 20 per cent. Above that, the data is weighted to correct for incompleteness.
\par
The passive galaxies in our groups exhibit a steep mass function, with no sign of a turnover above our limiting stellar mass of $M_{\rm star}=10^{10.3}M_\odot$. Apart from the lowest-mass bin, our data is in excellent agreement with the photometric analysis of \citet{2012A&A...538A.104G}, with the relative normalization based on the full sample shown in Figure~\ref{fig-mass_f_total}. This difference in the lowest-mass bin is also observed between the mass function in the \citet{2012arXiv1211.5607K} sample, which is based on an average of a large number of sparsely-sampled groups with a wide range of masses. Our work is complementary because we have a robust sample of groups with reasonably well-determined masses, but our statistics are not yet good enough to make a strong claim about this lowest mass point.
\par
In contrast with the groups, our field sample shows a lack of passive galaxies below $M_{\rm star}=10^{10.7}M_\odot$. This is reflecting the difference apparent in Figure~\ref{fig-cmass_z}, where the low-mass end of the red sequence of the field sample is apparently less well-populated than the group sample. This turnover is more prominent than in Giodini et al.'s field sample. The difference may be a cosmic variance effect, as the area of our survey is smaller, or it could reflect the fact that our field definition is systematically underdense because the dominant group in each pointing has been removed.
\par
Figure~\ref{fig-mass_f_blue} shows the stellar mass function for the blue (star-forming galaxies) in our group sample, again compared with the field and the \citet{2012A&A...538A.104G} data. The shape of the group mass function is very similar to that of the field, but with a normalization that is lower by a factor $\sim 2$. As these mass functions are normalized so the total mass functions match, this reflects a lower fraction of blue galaxies in our groups, at all stellar masses.
\subsubsection{The Fraction of Passive Galaxies}
\begin{figure}
	\leavevmode 
	\epsfysize=8cm
	\epsfbox{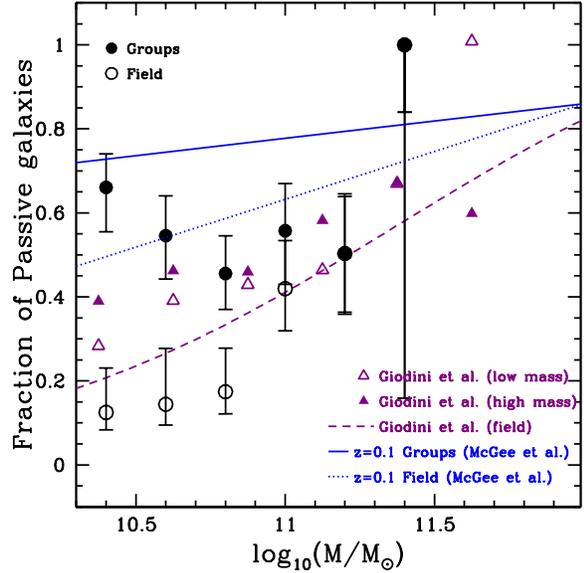} 
	\caption{The fraction of passive galaxies in both group and field samples is shown as a function of stellar mass. We compare these results to the photometric sample of \citet{2012A&A...538A.104G}, shown as the purple triangles with the same meaning as previous figures. The blue lines indicate the observed relations in the local universe as measured from the SDSS using a very similar method, from \citet{2011MNRAS.413..996M}. The solid line represents the local group population, while the dashed line is the field.}
	\label{fig-vred_f}
\end{figure}
\par
The steep mass function for passive galaxies leads to a large fraction of such galaxies at low masses.  We show this in Figure~\ref{fig-vred_f}, by dividing the passive-galaxy mass function by that for the full sample.   We compare this with the fraction in our spectroscopic field sample ({\it open symbols}), and the photometric field sample of \citet{2012A&A...538A.104G}. At high masses, the two are indistinguishable, though the uncertainties are large given our small volume. But at $M_{\rm star}<10^{11}M_\odot$ the groups remain dominated by purely passive galaxies, at $\sim 60$ per cent of the population, while the field is dominated by star-forming galaxies. Again, this reflects the absence of low-mass passive galaxies apparent in Figure~\ref{fig-cmass_z}.  
\par
Comparing next with the photometric group sample of \citet{2012A&A...538A.104G}, we see good agreement over most of the mass range. However, for $M_{\rm star}<10^{10.6}M_\odot$ we find the group population remains dominated by passive galaxies, where Giodini et al. note a continuing decline.  The origin of this discrepancy is not clear.  It possibly indicates a bias with the photometric method of Giodini et al., but as the most significant discrepancy is in a single bin, for which passive galaxies are only detectable in our lowest-redshift groups, this result should be tested with deeper spectroscopic data and a larger sample.
\par
As it stands, our data suggest that groups at $z\sim1$ were already dominated by passive galaxies at all stellar masses $M_{\rm star}>10^{10.3}M_\odot$. In \citet{2011MNRAS.413..996M} we presented an analysis of the SDSS and our GEEC groups at $z\sim0.5$, and identified passive galaxies based on their SED-derived SFRs. The results from that SDSS analysis are shown as the solid and dashed line for the group and field samples, respectively.  Our field sample clearly differs from that at $z\sim0.1$, with a steeper dependence of this fraction on stellar mass and an almost negligible passive population at the lowest masses. On the other hand, the passive fraction in the $z=1$ groups appears to be highest for the lowest masses, where the observed fraction is similar to that observed locally \citep[see also][]{2011ApJ...742..125G}. Again, much depends on the bin with the lowest stellar mass galaxies, and a deeper, larger spectroscopic sample is required to confirm this surprising result.  Certainly, we do not find any evidence that the fraction of passive group galaxies at a given stellar mass is ever lower than in the field. Using the \citet{2009ApJ...697.1842K} catalogue of the spectroscopic groups, \citet{2010A&A...509A..40I} have identified a mass-dependent environmental effect on the fraction of blue galaxies up to $z\sim0.8$, measured using their $(U-B)$ rest frame colour. This effect is strongest in the stellar mass range $M_{\rm star}<10^{10.6}M_\odot$, which is also the range where we observe the largest difference in the fraction of passive galaxies between the group and field. Using the local density estimator for the zCOSMOS sample, \citet{2010A&A...524A...2C} have also found that the local density modulates galaxy colour for this intermediate mass range ($10^{10.2}M_\odot<M_{\rm star}<10^{10.7}M_\odot$). Our spectroscopically complete sample demonstrates this density dependence was in place at $z\sim1$, corresponding to a look back time up to 1 Gyr earlier than the maximum probed in these studies.
\subsection{Spectroscopic Properties of the Red, Blue, and Green Populations}
\par
\begin{figure}
	\leavevmode
	\epsfysize=8cm
	\epsfbox{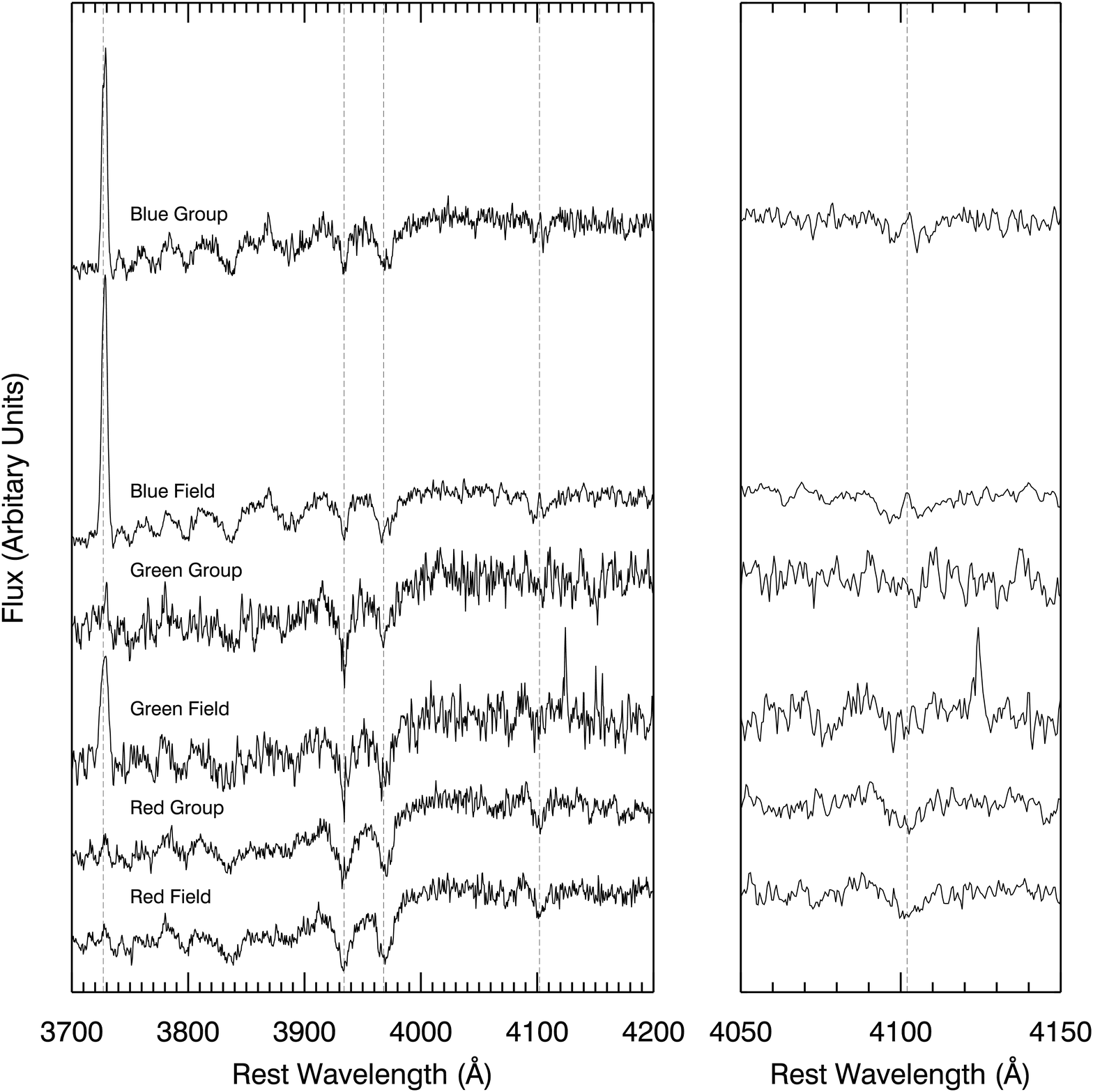} 
	\caption{Rest-wavelength stacked spectra for galaxies at $0.8<z<1$ and $9.5<\log (M_\mathrm{star}/M_\odot)<11.6$, including the spectroscopic and mass-completeness weights, for the six main classification used in this paper. The spectra are arbitrarily normalized within the wavelength range of $4050-4100$ \AA. The location of the important lines (in order: \oii, Calcium H\&K, and H$\delta$) are also indicated on this plot. The figure on the right show an expanded section of the spectra around the \hdelta line.}
	\label{S01_Spectra}
\end{figure}
\par
Next, it is instructive to consider the average properties of each population. Figure~\ref{S01_Spectra} shows an example of the stacked spectra for the six sub-populations, with relatively high signal-to-noise ratios of between $7.0$ per \AA\ (in the range of 4050 \AA\ to 4100 \AA) for the green field stacked spectra to $12.7$ per \AA\ for the more numerous blue group stacked spectra.
\par
The average spectra confirm that green galaxies have features that are intermediate between blue and red galaxies. They have less prominent absorption lines, and do not resemble typical post-starburst type galaxies. Second, \oii emission is present at a low level, much more so in the field population. Thus, some star formation is still ongoing in these galaxies. 
\par
Using the stacked spectra, the ``average'' spectral properties for the red, blue, and green populations can be determined by measuring the \oii and \hdelta strengths. Note that our stacked spectra contain enough signal to distinguish the presence of weak emission-filling in the \hdelta absorption line of blue galaxies. We correct for this in the calculation of the average \hdelta equivalent width, by fitting multiple Gaussian functions. The equivalent width of the emission line is then added to the total equivalent width obtained using the automated bandpass method, as described in \S~\ref{sec-spec}.\footnote{Note that for the lower mass blue galaxies bin, the emission component was measured to be 1.5 \AA\ for the group galaxies and 1.3 \AA\ for the field galaxies. For the higher mass bin, the emission component was measured to be $\sim1.8$ \AA\ for the group and field population.} For the blue population, we separated the populations into a low mass and high mass sample at $\mathrm{log}(M_\mathrm{star}/M_\odot)=10.3$. This decision was motivated by the differences observed in the distribution of sSFR in \S~\ref{sec-oii_ssfr_m} between the high mass and low mass blue galaxies, caused in part by the onset of incompleteness.
\par
\begin{figure}
	\leavevmode
	\epsfysize=8cm
	\epsfbox{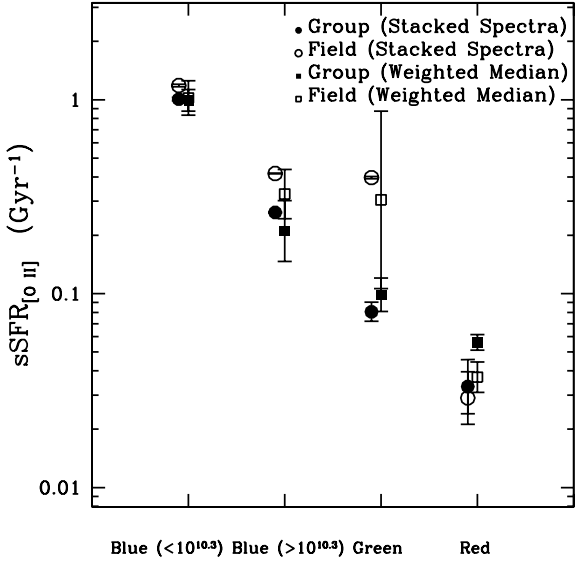} 
	\caption{The sSFR for the red, green, and blue populations. The galaxies are restricted to $0.8<z<1$ and include galaxies within the GMOS fields-of-view and total weights (spectroscopic and mass-completeness) of between 0 and 5. Bins of group galaxies are shown with {\it filled symbols} and field galaxies as {\it open symbols}. The red and green bins contain galaxies with masses $9.5<\log (M_\mathrm{star}/M_\odot)<11.6$. For the blue population, another cut was made, separating the two populations at $\mathrm{log}(M_\mathrm{star}/M_\odot) = 10.3$. The {\it circle symbols} indicate the results from stacked-spectra measurement of \oii sSFR. The {\it squares symbols} are data from the same bin, except combined using the weighted median method.}
	\label{C05_sSFR}
\end{figure}

First, the sSFR for these four bins are shown in Figure~\ref{C05_sSFR}, with filled and open circles for the group and field populations, respectively. For the red and blue populations, this average shows at most a weak dependence on environment. For the blue galaxies in particular, where the statistical uncertainty on the stacked spectra is negligible, we observe a significant difference between the average sSFR of the low and high mass blue sample. While the group galaxies have only $\sim3$ per cent lower sSFR than the field sample for the low mass blue population, the difference is $\sim55$ per cent for the high mass sample.  The apparently high significance of this difference is misleading, however, as the error bar only reflects the fact that the \oii feature is measured with high signal-to-noise in the stacked spectrum.  Intrinsic variations in the line strength amongst galaxies means the average SFR is not really so well determined, as already  seen in Figures~\ref{BG05_sSFR} and \ref{Histo_Res_LM}.
\par
Interestingly, the intermediate, green galaxies show an even larger difference, as green group galaxies have an average sSFR a factor $\sim3$ lower than the field, when measured using the stacked spectra. This result does need to be treated carefully, as this is a relatively small population defined by sharp boundaries amongst a continuous distribution (Figure~\ref{fig-ccproj}). Both statistical and systematic (i.e. the connection between colours and underlying star formation history) uncertainties mean that the green population likely contains ``contamination'' from the adjacent blue and red populations. Since the red fraction is a strong function of environment, this can lead to an apparent environmental dependence amongst these green galaxies.
\par
When we compare their sSFRs from the weighted-median of the individual galaxy measurements (see \S~\ref{sec-spec}), shown as the squares in Figure~\ref{C05_sSFR}, the difference between group and field still exists. However, the error bars now include the bootstrap contribution to the error due to the scatter of the parent distribution. The larger error bars indicate that these differences in the sSFR are not statistically significant, given the small sample sizes and the wide spread in the properties of the binned galaxies. Therefore, we cannot confidently claim to detect any significant suppression of sSFR amongst group galaxies of a given spectral type, relative to the field.
\begin{figure}
	\leavevmode
	\epsfysize=8cm
	\epsfbox{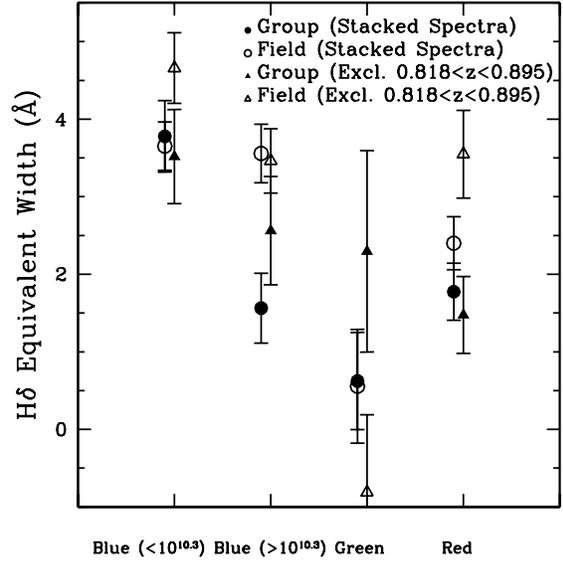} 
	\caption{The H$\delta$ equivalent widths for the red, green, and blue populations. The galaxies are restricted to $0.8<z<1$ and include galaxies within the GMOS fields-of-view and total weights (spectroscopic and mass-completeness) of between 0 and 5. Bins of group galaxies are shown with {\it filled symbols} and field galaxies as {\it open symbols}. The red and green bins contain galaxies with masses $9.5<\log (M_\mathrm{star}/M_\odot)<11.6$. For the blue population, another cut was made, separating the two populations at $\mathrm{log}(M_\mathrm{star}/M_\odot) = 10.3$. In addition, the measurement of the blue galaxies includes a correction for \hdelta emission, computed using a Gaussian fit to the stacked spectra. The {\it triangles} represents the removal of galaxies in the redshift range $0.818<z<0.895$, in order to avoid the telluric-dominated spectra.}
	\label{C02_HDelta}
\end{figure}
\par
Next, we consider the H$\delta$ absorption line. This spectral feature is most prominent in A-type stars and is a signature of recent star formation, especially for galaxies with low instantaneous sSFR. An especially strong H$\delta$ signature, with an equivalent width greater than 5 \AA, may indicate a truncation in star formation, due to the domination of intermediate age A-type stars \citep[e.g.][]{1987MNRAS.229..423C, 1999ApJ...527...54B, 2012ApJ...745..106L}.
\par
In Figure~\ref{C02_HDelta}, we show the H$\delta$ equivalent width measured using the stacked spectra as the circular points. One potential source of concern is that the corrections to the telluric absorption at the A-band may not be perfect. They often leave significant residuals which unfortunately overlap with H$\delta$ over the redshift range, $0.818<z<0.895$. To ensure that this contamination is not affecting our results, we remove galaxies in the redshift range $0.818<z<0.895$ and show our updated results for this restricted sample with triangular points.
\par
We find differing environmental signatures in \hdelta between the red, green, and blue populations, though they are not highly significant given our uncertainties. The blue galaxies ($M_{\rm star}/M_\odot>10^{10.3}$) show a weaker absorption for the group sample, as compared to the field, which may be related to the observed differences in their sSFR from Figure~\ref{C05_sSFR}. However, this significant difference disappears when we consider the sample unaffected by telluric absorption, so we again treat this result with caution. Another result to note from the \hdelta measurements is that the green population is not dominated by post-starburst galaxies (\ewhdelta$>5$ \AA), as observed from their spectra. Surprisingly, the \hdelta strength of the green population is much less than that of the blue population, which suggests that the last period of significant star formation occurred $\gtrsim1-2$ Gyr ago. In addition, the green field galaxies have a weaker \hdelta line than even the red galaxies, though this result should be treated with caution. The systematic errors are not included in the error bars, as demonstrated by the substantial shift after the removal of the telluric dominated galaxies. Furthermore, these green galaxies may have an undetected \hdelta emission component in their low signal-to-noise spectrum, especially for the field population in which we see significant \oii emission.
%
%
\section{Discussion}\label{sec-discuss}
\par
\begin{figure}
	\leavevmode
	\epsfysize=8cm
	\epsfbox{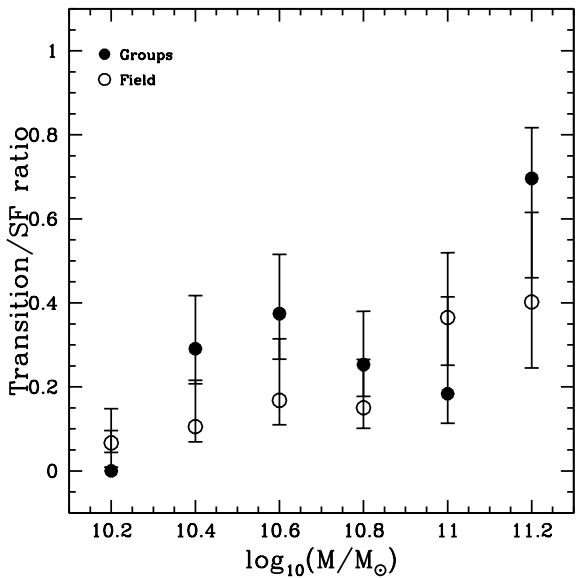} 
	\caption{The ratio of the mass functions of the transition population and the star-forming population are shown, for our group ({\it filled circles}) and field samples ({\it open circles}). The ratio increases with stellar mass, and shows little environmental dependence. \label{fig-green_f}}
\end{figure}

We proposed in \citet{2011MNRAS.412.2303B} that the intermediate green population may represent a transition from the blue cloud to red sequence \citep[see also][]{2008MNRAS.385.2049G, 2012ApJ...745..106L}. In Figure~\ref{fig-green_f} we show the ratio of the ``green'' and ``blue'' mass functions. If the former population is indeed a transition from the latter, this ratio should give some indication of the rate at which this transformation is occurring. What we see is that this fraction is largest for the highest mass galaxies, and is $\sim 20$ per cent at $M_{\rm star}<10^{10.7}M_\odot$.  There is no strong evidence that the fraction is any higher within groups, though the uncertainties are large enough to be consistent with an interestingly-large difference. If the green galaxies do represent a transition population, many of them may be more closely associated with the mass-quenching population \citep{2010ApJ...721..193P}. Another possibility is that some of them represent predominantly passive galaxies that have been partially rejuvenated, for example following a minor merger event which provides fuel for additional star formation.  On the other hand, many of the field, green galaxies may be satellites of smaller groups or large galaxies.  Thus the possibility remains that this population is largely environment-driven, a result that can be tested with highly complete spectroscopic surveys of the field at this redshift.
\par
\begin{figure}
	\leavevmode
	\epsfysize=8cm
	\epsfbox{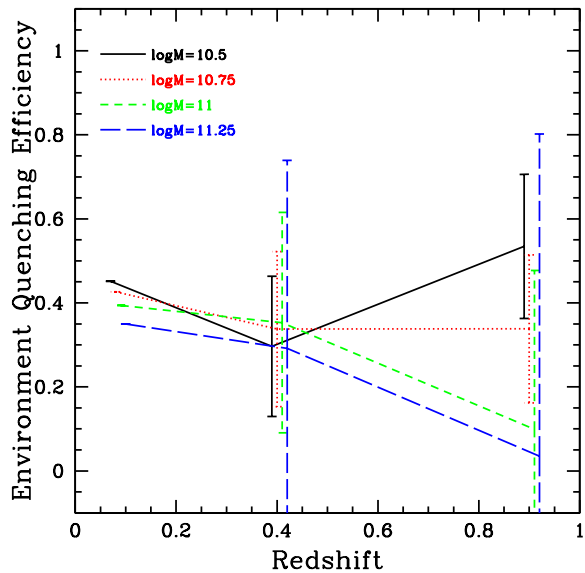} 
	\caption{The environment quenching efficiency, $\epsilon_\rho$, as defined by \citet{2010ApJ...721..193P}, is shown for four different stellar mass bins and three redshifts. The $z=0$ data are based on the SDSS; these and the $z=0.5$ data are taken from the measurements in \citet{2011MNRAS.413..996M}, interpolating to the mass bins used here.}
	\label{fig-red_f_z}
\end{figure}

Despite the fact that most of the transition population do not appear to be related to the group environment, the fact that we observe significantly higher fractions of passive galaxies in the groups than the field, for $M_{\rm star}<10^{10.7}M_\odot$, implies that environment does play a role. Assuming that the properties of the observed field galaxies are similar to the pre-accreted population, we can compute an efficiency factor that describes the fraction of field galaxies at a given mass that would have to be quenched to match the red fraction at a given over-density.  Following the spirit of \citet{2010ApJ...721..193P}, we define this environment quenching efficiency at stellar mass $M_{\rm star}$, as:
\begin{equation}\label{eqn-erho}
	\epsilon_\rho = \frac{f_{\rm red}(\mbox{group},M_{\rm star})-f_{\rm red}(\mbox{field},M_{\rm star})}{f_{\rm blue}(\mbox{field},M_{\rm star})}.
\end{equation}
 Similar definitions are advocated by \citet{2008MNRAS.387...79V} and \citet{2012arXiv1206.3571W}.  Using the results from \citet{2011MNRAS.413..996M} we compute $\epsilon_\rho$ at $z=0$ and $z=0.5$, and compare this with our new data, in Figure~\ref{fig-red_f_z}. While the uncertainties are still fairly large, a value of $\epsilon_\rho\sim 0.4$ is consistent with our data and is generally independent of both stellar mass and redshift. The value for the environmental quenching efficiency is also consistent with similar analysis done up to $z\sim0.8$ from the larger zCOSMOS spectroscopic group catalogue \citep{2012arXiv1211.5607K}. There are hints of a smaller $\epsilon_\rho$ for the higher masses, though the uncertainties are large.
\begin{figure}
	\leavevmode
	\epsfysize=6cm
	\epsfbox{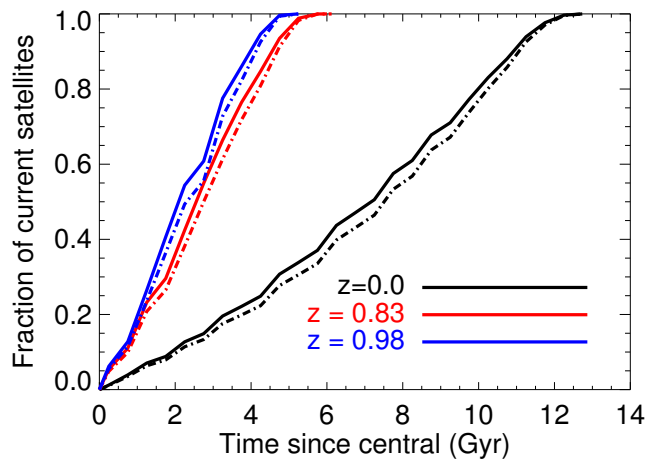}
	\caption{From the galaxy formation model of \citet{2006MNRAS.370..645B}, we show the cumulative fraction of satellites in groups of the indicated masses and at the indicated redshifts ($z=0.0$, $z=0.83$, $z=0.98$), as a function of the time since they were last identified as a ``central'' galaxy. The solid line indicates accretions into groups with log($M_\mathrm{grp}/M_\odot)>13.0$ and dashed line indicates groups with log($M_\mathrm{grp}/M_\odot)>13.5$. The model gives an indication of how long satellite galaxies may have been sensitive to environment-quenching. The stellar mass cut-off for the satellites is set at $>10^{10}M_\odot$.}
	\label{fig-accret}
\end{figure} 
\par
\citet{2012arXiv1206.3571W} recently argued, based on the shape of the SFR distribution amongst satellite galaxies in the SDSS, that the shutdown of star formation in satellites is best characterized by a long ($\sim 3$ Gyr), mass-dependent delay where evolution continues as normal, followed by rapid ($\tau_q \lesssim 1$ Gyr) quenching. If this timescale does not evolve with redshift, we might therefore expect that star-forming galaxies accreted between $3$ Gyr and $3+\tau_q$ Gyr ago are in the process of quenching at the epoch of observation.
\par
However, the long delay, also advocated by \citet{2012MNRAS.423.1277D}, may be difficult to reconcile with the large environmental effect we observe. The galaxy formation models of \citet{2006MNRAS.370..645B} predict the fraction of satellites accreted for groups of our mass and redshift as a function of time, as shown in Figure~\ref{fig-accret}. The results show that the fraction accreted in time $\Delta \tau$ is $\sim 20 \Delta \tau/\mbox{ Gyr}$ per cent of the total observed population. Less than $\sim 30$ per cent of satellites are predicted to have been a satellite for as long as 3.5 Gyr before the epoch of observation. Thus it is difficult to see how an environmental signature as strong as that seen in Figure~\ref{fig-vred_f} could arise. The longer timescale of $\sim 7$ Gyr advocated by \citet{2012MNRAS.423.1277D} is even more problematic, as this is longer than the age of the Universe at $z=1$. If the ``delayed-then-rapid'' scenario of \citet{2012arXiv1206.3571W} is correct, then it seems the delay must be considerably shorter at $z=1$. For example, if it scales with the dynamical time as $\tau\propto (1+z)^{-3/2}$ \citep{2010ApJ...719...88T}, we might expect the relevant timescale at $z=1$ to be $1$--$2$ Gyr. This would likely be consistent with our observations.
\par
In principle we can use the observed fraction of transition galaxies to try and constrain the quenching time, $\tau_q$. We find that green galaxies make up $10$ per cent of the total galaxy population, at $M_{\rm star}<10^{10.7}M_\odot$. This must be an upper limit to the fraction undergoing transition at the observed epoch, as at least some are likely to result from mass-quenching, or rejuvenated star formation in old galaxies. Given the high accretion rates shown in Figure~\ref{fig-accret}, the quenching time must be quite short, $\tau_q\sim 0.5$ Gyr or less, in order to avoid overproducing galaxies in this phase of evolution. In a future paper, we will use a simple model, using Figure~\ref{fig-accret}, to predict the red fractions and the expected H$\delta$ strengths of the blue, green, and red populations using population synthesis models. This will allow us to constrain $\tau_q$ and the delay timescale, using our observations.
%
%
\section{Conclusions}
\par
We have presented highly complete GMOS-S spectroscopy for galaxies with $r<24.75$ in the fields of 11 galaxy groups at $0.8<z<1$. We use an optical-NIR colour-colour diagnostic diagram to divide the galaxies into three subpopulations: star-forming (blue), passive (red), and intermediate (green). Using SED-fitting techniques and spectral analysis, we present a comparison of group populations with the field, down to a stellar mass of $M_{\rm star}>10^{9.5}M_\odot$. We conclude the following:
\begin{itemize}
\item The strongest environmental dependence is on the fraction of passive galaxies. For galaxies with $10^{10.3}<M_{\rm star}/M_\odot<10^{11}$, groups are dominated by passive galaxies, which only constitute $\sim 20$ per cent of the field at these stellar masses.
\item For the normal star-forming galaxies, the average sSFR for the high mass ($M_{\rm star}/M_\odot>10^{10.3}$) group sample is $\sim55$ per cent lower than the field population, but the difference is not statistically significant. For the low mass blue galaxies, where our sample is incomplete, we observe no difference in sSFR between the group and the field.
\item The green galaxy population represents approximately the same fraction of the star-forming population at all stellar masses, in both the group and the field. Neither the group or field spectra show very strong \hdelta absorption, which indicates that their populations are not dominated by post-starburst galaxies. Green galaxies in the group exhibit a factor of $\sim3$ lower sSFR than the field, calculated using their stacked spectra, suggesting that the origins of these transitional systems may be environment dependent. 
\end{itemize}
\par
In the context of the simple model proposed by \citet{2010ApJ...721..193P}, our results are consistent with an environment-quenching efficiency of $\sim 40$ per cent (for $z\lesssim1$). There are hints that this value may be smaller for galaxies with higher stellar masses. Furthermore, the abundance of possible ``transition'' galaxies requires a relatively short quenching timescale of $\tau\lesssim 0.5$ Gyr. Various authors have argued there must be a long timescale, on the order of $\sim 3-7$ Gyr, associated with satellite quenching \citep{2009MNRAS.400..937M, 2012MNRAS.423.1277D, 2012arXiv1206.3571W}. This could be associated with a delay, before star formation is truncated on a shorter timescale. However, such a long delay time seems incompatible with the large environmental signature we see at $z=1$, where most galaxies have not spent that long as satellites. Either this timescale must evolve so that it is shorter at higher redshift \citep{2010ApJ...719...88T}, or perhaps only a fraction of galaxies ever respond to their environment. Better constraints could be obtained by extending this analysis to lower-mass galaxies with low or negligible star formation, which will require deep integrations of NIR-selected samples.
%
%
\section{Acknowledgments}
\par
We would like to thank the referee for their helpful comments. We are grateful to the COSMOS and zCOSMOS teams for making their excellent data products publicly available. This research is supported by NSERC Discovery grants to MLB and LCP. We thank the DEEP2 team, and Renbin Yan in particular for providing the {\sc zspec} software, and David Gilbank for helping us adapt this to our GMOS data. Finally, we appreciate thoughtful comments on the applicability of different SFR indicators, from Felicia Ziparo.
%
%
\bibliographystyle{mn2e}
\bibliography{report}
%
%
\appendix
\section{Normalization Difference between Dust-Corrected \oii SFR and FUV+IR SFR}\label{app1}
\par
To calculate the FUV+IR SFR, we first k-correct the UV and optical photometric data from the zCOSMOS catalogue using {\sc kcorrect} {\sc IDL} software of \citet{2007AJ....133..734B} to their rest frame. To do this, we apply corrections to the filter that most closely matches the rest wavelength of the desired filter, rather than taking the synthetic magnitude from the template fit.  This allows us obtain reasonable upper limits for galaxies that are undetected in rest-frame FUV flux. Next, MIPS 24 micron data is converted into total IR luminosity with the templates from \citet{2001ApJ...556..562C}. We then combine FUV and IR luminosities into SFR estimates with the prescription of \citet{2011ApJ...741..124H}. Finally, galaxies without MIPS 24 micron data had dust attenuation estimated using the ${({\rm FUV-NUV})^0}$ relationship from \citet{2009ApJ...700..161S}.
\par
\begin{figure}
	\leavevmode
	\epsfysize=8cm
	\epsfbox{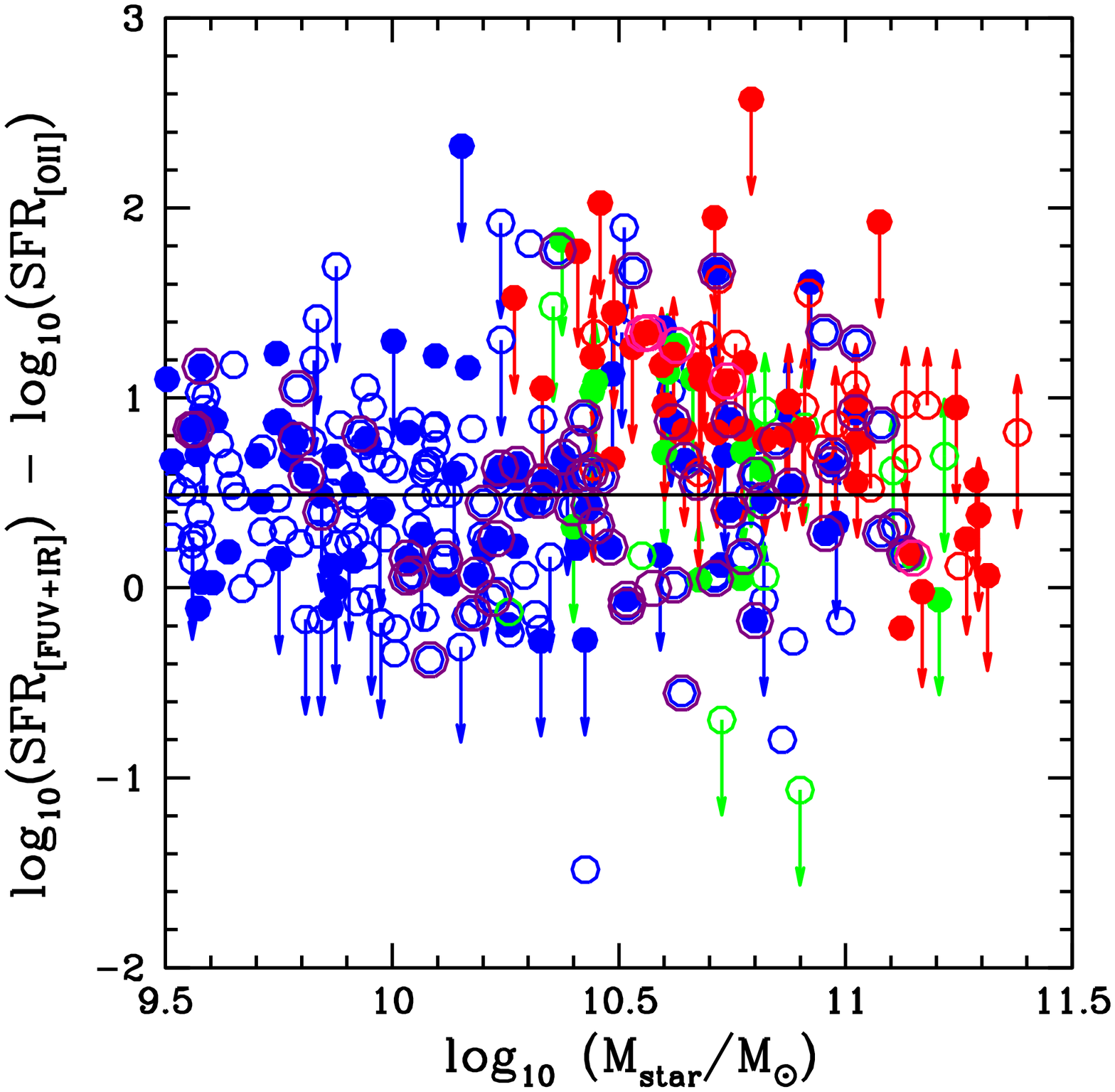} 
	\caption{Log$_{10}$ (SFR$_{\rm[FUV+IR]}$) - log$_{10}$ (SFR$_{\rm[OII]}$) is plotted as a function of stellar mass for galaxies at $0.8<z<1$ in the GMOS fields-of-view. Group members are shown as {\it filled symbols}, and field members within our GMOS fields-of-view are shown as {\it open symbols}. Only galaxies with total weight (spectroscopic and mass-completeness) of between 1 and 5 are included. Points are colour-coded according to their classification as red, blue, and green. Galaxies with 24 micron MIPS detection within 3$''$ are outlined in purple. The black line indicates the mass-independent offset between the FUV + IR and OII star formation rates. Note that those galaxies with a negative \oii equivalent width or with \oii detections less than the measurement error are converted into lower limits, while undetected FUV galaxies are shown as upper limits.}
	\label{Comp_SFR_M_FUVIR3S_OII}
\end{figure}

In Figure~\ref{Comp_SFR_M_FUVIR3S_OII}, the log difference in the SFR between the two methods, log$_{10}$ (SFR$_{\rm[FUV+IR]}$) - log$_{10}$ (SFR$_{\rm[OII]}$), is shown as a function of stellar mass. A comparison of the \oii dust corrected SFR and the FUV + IR SFR for our sample show a normalization offset, as noted in \citep[e.g.][]{2011ApJ...735...53P, 2011ApJ...730...61K}. A mass-independent difference is observed, and so the \oii - derived SFR are multiplied by a factor of 3.1 for the purpose of this paper.  This will be explored in more detail in a future paper (Mok et al., in prep.).
%
%
\end{document}